\documentclass[9pt,twocolumn,twoside]{pnas-new}
\usepackage{soul}
\usepackage{verbatim}
\templatetype{pnasresearcharticle} 

\title{Learning-dependent navigation strategies and neural computations in \emph{Caenorhabditis elegans}}

\author[a]{Kevin S. Chen}
\author[b]{Anuj K. Sharma}
\author[a]{Jonathan W. Pillow\textsuperscript{1,2}} 
\author[a,b]{Andrew M. Leifer\textsuperscript{1,2}}

\affil[a]{Princeton Neuroscience Institute, Princeton University, Princeton, NJ 08544}
\affil[b]{Department of Physics, Princeton University, Princeton NJ, 08544}

\leadauthor{Chen} 

\significancestatement{
Learning is a feature of species across scales. How does the brain flexibly learn to produce complex behavior? We focus on \emph{C. elegans} to study how navigation strategies depend on olfactory learning by utilizing precise measurements and a novel model. The fitted model shows that learning alters two known navigation strategies in worms and the model outperforms a classical metric in decoding the animal's past learning experience. We discover that learned navigation can express itself in a context-dependent manner. 
In addition, through perturbing individual interneurons, we find that most neurons' contribution are distributed across strategies and are differentially altered by learning. We expect these flexible behavioral algorithms and neural computation to be generalized to other species and behavior.}

\authorcontributions{K.S.C, J.W.P., and A.M.L designed the research; K.S.C. performed research, analyzed data, and wrote the original draft; A.K.S. contributed new reagents; K.S.C, J.W.P., and A.M.L wrote the paper.}

\authordeclaration{The authors declare no conflict of interest.}
\equalauthors{\textsuperscript{1}J.W.P. and A.M.L contributed equally to this work.}
\correspondingauthor{\textsuperscript{2}To whom correspondence should be addressed. E-mail: leifer@princeton.edu, pillow@princeton.edu}


\begin{abstract}
Learned olfactory navigation is a powerful platform for studying how a brain implements associative learning of complex behaviors. However, the quantitative change in sensorimotor transformation and the underlying neural circuit substrates to generate learned goal-directed navigation are still unclear.
Here we investigate learning-dependent sensorimotor processing and the neural basis for navigation in the nematode \textit{Caenorhabditis elegans} by measuring, modeling, and perturbing learned odor-guided navigation. We develop a novel statistical model to characterize how the worm employs two behavioral strategies: a biased random walk and weathervaning. We infer weights on these strategies and characterize sensorimotor kernels that govern them by fitting our model to the worm's time-varying navigation trajectories and precise odor concentration measurements. Compared to naive worms, appetitive trained worms up-regulate their biased random walk strategy, while aversive trained worms down-regulate their weathervaning strategy. The model predicts an animal's past learned experience with $>90 \%$ accuracy given finite observations, outperforming a classical chemotaxis metric. The model trained on natural odors further predicts the animals' learning-dependent response to optogenetically induced odor perception. Our measurements and model show that behavioral variability is altered by learning--- trained worms exhibit less variable navigation than naive ones. Genetically disrupting individual interneuron classes downstream of an odor-sensing neuron reveals that learned navigation strategies are distributed in the network. Together, we present a flexible navigation algorithm that is supported by distributed neural computation in a compact brain.

\end{abstract}

\dates{This manuscript was compiled on \today}
\doi{\url{www.pnas.org/cgi/doi/10.1073/pnas.XXXXXXXXXX}}

\begin{document}

\maketitle
\thispagestyle{firststyle}
\ifthenelse{\boolean{shortarticle}}{\ifthenelse{\boolean{singlecolumn}}{\abscontentformatted}{\abscontent}}{}


\dropcap{L}earning is a fundamental property of neural systems that allows an animal to flexibly alter behavior based on experience. 
Learned olfactory navigation provides an ideal framework for studying learning because it is a naturalistic behavior, common to species across scales \cite{baker2018algorithms, porter2007mechanisms}, ethologically relevant for seeking food \cite{Torayama2007-qi, gire2016mice} and avoiding pathogens \cite{Zhang2005-if, Ha2010-ja}, and can be rapidly learned in very few trials \cite{gire2016mice,krakauer2017neuroscience, meister2022learning}. Careful characterization of learned navigation behavior in a naturalistic context can shed light on the flexible computation performed by the brain \cite{clark2013mapping}.

To study how animals flexibly alter their olfactory navigation upon learning, we focus on the nematode worm \emph{C. elegans}. This worm has a compact nervous system \cite{white1986structure}, well-characterized olfactory neural circuits \cite{Bargmann2006-dy, Pritz2023-du, Cook2019-dl, dusenbery1975chemotaxis, croll1977sensory} and navigation behavior that has been studied in detail \cite{Bargmann2006-dy, Pierce-Shimomura1999-nt}. Worms learn to associate an odor with the presence or absence of food and then will either navigate towards higher concentrations of the odor or will ignore the odor, respectively \cite{
Torayama2007-qi, Kauffman2011-up}. However, it is unknown how the worms' navigation strategies are altered by learning to achieve these feats. We aim to quantitatively characterize the learned navigation strategy and their underlying sensory transformations to constrain neural mechanisms of learned navigation.

The worm navigates in sensory environments mainly through two strategies: klinotaxis and klinokinesis \cite{Pierce-Shimomura1999-nt, Iino2009-al, Ikeda2020-tw}. Klinotaxis is a process in which the worm continuously modulates its heading to align with the local gradient, also known as ``weathervaning'' \cite{Iino2009-al}. Klinokinesis is a biased random walk \cite{macnab1972gradient, keller1971model} in which the worm produces sharp turns called ``pirouettes'' with a probability that depends on the animal's estimate of the local gradient of a sensory cue \cite{Pierce-Shimomura1999-nt, berg1993random, macnab1972gradient}. In both cases the animal estimates information about the local gradient by comparing measurements of the stimuli across time. Both strategies contribute to sensory-guided navigation in landscapes of temperature \cite{Ikeda2020-tw}, salt \cite{Iino2009-al, Luo2014-pc, Jang2019-np}, and certain odors \cite{Iino2009-al, Bargmann2006-dy, Colbert1995-vh, Tanimoto2017-pt}. 

Olfactory navigation has significant relevance in studying learning because there are odors for which the animal learns to alter its ``valence'' (positive or negative) with respect to the odor--- similar to  associative learning in many other animals. This contrasts with context-dependent salt or thermosensing in \textit{C. elegans} in which the worm instead learns a preferred salt concentration or temperature set point based on past experience \cite{Luo2014-pc, Ikeda2020-tw}. For learned olfactory navigation, it is not known how the biased random walk and weathervaning strategies change upon learning, nor how the detailed quantitative transformations between sensory input and motor output are changed. 

Quantitative analysis for airborne odor-guided navigation has been limited, in part because of experimental challenges in measuring detailed information about the odor concentration experienced by the animal. While past work estimated odor cues indirectly or with models  \cite{gire2016mice, Iino2009-al}, precise concentration measurements are required to characterize sensorimotor transformation in the sensory environment. Recently, however, new experimental methods to control and monitor the odor landscape experienced by small animals, such as the worm, \cite{Chen2023-fy, tadres2022depolarization, gershow2012controlling}, now make it possible to empirically constrain quantitative models of odor-guided navigation, as we pursue here.

Worms are intrinsically attracted to butanone, a volatile organic compound found in bacterial food in the worm's natural habitat \cite{Worthy2018-xi}. When worms are exposed to butanone paired with food (``appetitive training''), they increase their attraction and are more likely to navigate towards higher butanone concentrations \cite{Torayama2007-qi, Kauffman2011-up, Cho2016-is, Pritz2023-du}. In contrast, when butanone is paired with starvation (``aversive training''), worms decrease their tendency to climb up butanone gradients in comparison to worms without exposure to butanone (``naive'') \cite{Torayama2007-qi, Cho2016-is}. The animal's neural and behavioral responses to butanone both change upon learning \cite{Cho2016-is, Pritz2023-du}. Sensory neuron AWC$^\textrm{ON}$ responds to butanone \cite{Cho2016-is, Torayama2007-qi}, as well as others \cite{Lin2023-gf, Pritz2023-du}. However, the involvement of downstream interneurons in butanone learning and learned navigation are still unclear.

In this study we seek to answer: (1) How are the worm's navigation strategies altered by olfactory learning?  (2) How are sensorimotor transformations altered by learning and how does this vary across a population? (3) What neural substrates may be involved? To answer these questions, we combine precise experimental measurements using our recently developed continuous odor monitoring assay \cite{Chen2023-fy} and a novel statistical model to rigorously characterize how butanone associative learning alters odor navigation strategies in worms.  

Our measurements and model reveals that the animal's biased random walk is bidirectionally altered by butanone learning and that its weathervaning strategy is down-regulated upon aversive learning. Our approach yields interpretable model parameters that better decode training conditions compared to chemotaxis index, and also predict response to optogenetic perturbation in the sensory neuron AWC$^\mathrm{ON}$. We discover that naive worms have higher behavioral variability and demonstrate context dependent behavior. And we provide insights into the role of specific interneurons. 

\begin{figure*}[t]
\centering
\includegraphics[width=1.\linewidth]{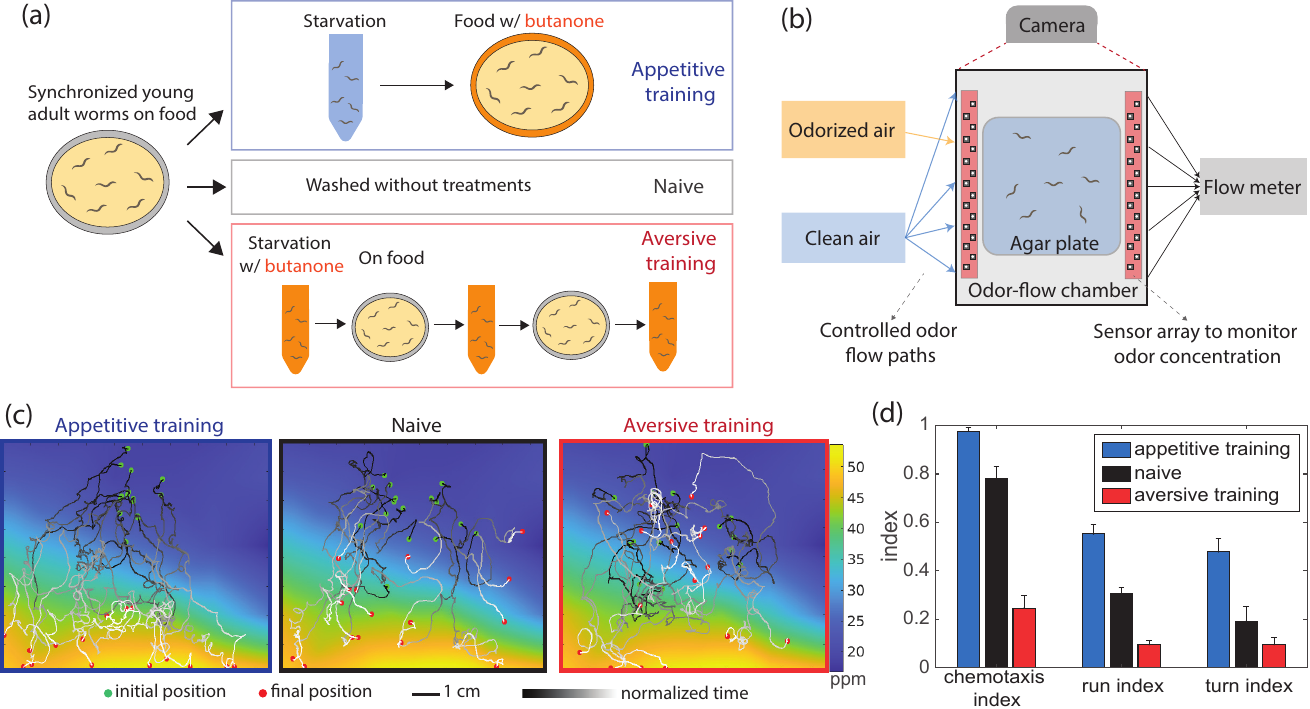}
\caption{Bidirectional olfactory learning in \emph{C. elegans}. \textbf{(a)}  Protocol for butanone associative training in worms. \textbf{(b)} After exposure to different training regimens in (a), worms'  olfactory navigation is measured in a controlled odor environment. \textbf{(c)} Example trajectory after three different training conditions. \textbf{(d)} Summary statistics of learning across three conditions. The chemotaxis index is the normalized number of tracks going up gradient $N_u$ versus down $N_d$: $\frac{N_u-N_d}{N_u+N_d}$, which corresponds to the classic chemotaxis index. The run index is the normalized run length going up gradient $r_u$ versus down $r_d$: $\frac{r_u-r_d}{r_u+r_d}$, and captures a key feature of a biased random walk. The turn index is the normalized probability turning up gradient $p_u$ versus down $p_d$: $\frac{p_u-p_d}{p_u+p_d}$, which captures a key feature of weathervaning. Error bar shows standard error of mean across 7-13 plates for each condition. For each index, all pairs of training conditions (appetitive-naive, naive-aversive, and appetitive-aversive) show statistically significant differences (t-test, $p<0.05$), except for the turn index when comparing naive to aversive conditions.
}
\label{fig:fig1}
\end{figure*}

\section*{Results}

\subsection*{Learning bidirectionally alters olfactory navigation}
We developed a protocol to train worms to associate butanone with either food (appetitive training) or starvation (aversive training) (Figure \ref{fig:fig1}a). Our protocol was similar to previously reported training regimens \cite{Kauffman2011-up, Cho2016-is, Pritz2023-du} except that ours exposes the animal to multiple rounds of odor paired with starvation instead of only one, which we found increases consistency in learning (Fig. S2).  
After training, we recorded the movement of populations of worms as they crawled in a defined odor landscape that used metal-oxide sensors to continuously monitor the odor concentration along the boundary \cite{Chen2023-fy} (Figure \ref{fig:fig1}b, Fig. S1). We recorded hundreds of locomotory trajectories per plate in this odor environment after different training conditions, for up to 13 plates per training condition.  Animal's locomotory trajectories were qualitatively different depending upon learning, with appetitive trained animals traveling up gradient more often than naive animals, and aversive-trained animals traveling up gradient least of all, broadly consistent with prior reports \cite{Cho2016-is, Torayama2007-qi, Pritz2023-du} (Figure \ref{fig:fig1}c). We quantify the performance of traveling up the gradient using a ``chemotaxis index'' that is calculated by comparing the number of trajectories going up-gradient versus down gradient \cite{Cho2016-is, Torayama2007-qi} (Figure \ref{fig:fig1}d). 
Performance navigating up gradients is bidirectionally modulated by learning: chemotaxis index increases after appetitive training and  decreases after aversive training compared to naive worms that undergo no training. Interestingly, even animals that undergo aversive training do not, on average, navigate  down the gradient, suggesting that after learning an association between butanone and starvation, animal are indifferent or possibly still slightly attracted to butanone.

To explore whether worm navigation superficially resembles a  
biased random walk, we calculated a ``run index'' by computing the normalized length of a run moving up-gradient, where the run is defined as the period between pirouettes. To explore whether navigation superficially resembles a weathervaning strategy, we calculated a ``turn index'' that reports the normalized fraction of turn events that result in heading up-gradient \cite{Luo2014-pc}. Our measurements show that both the run index and the turn index are on average bidirectionally modulated by learning with respect to naive animals, suggesting that learning alters both of these navigational strategies. 

While the metrics above suggest hypotheses about how navigational strategies change due to learning, they provide little information about the dynamics of navigation, nor the sensorimotor transformations that govern these dynamics. Specifically, the metrics use only binary information about whether the animal is traveling up or down gradient, and ignore details of the sensory landscape like the odor concentration experienced over time. The indices also provide no information about the behavioral variability across the population. To overcome these limitations, we sought a statistical model that captures temporal information, behavioral noise, and explicitly predicts how the animal changes its movement in response to sensory stimuli. 

\subsection*{Odor-dependent mixture model of olfactory navigation}

\begin{figure*}
\centering
\includegraphics[width=1.\linewidth]{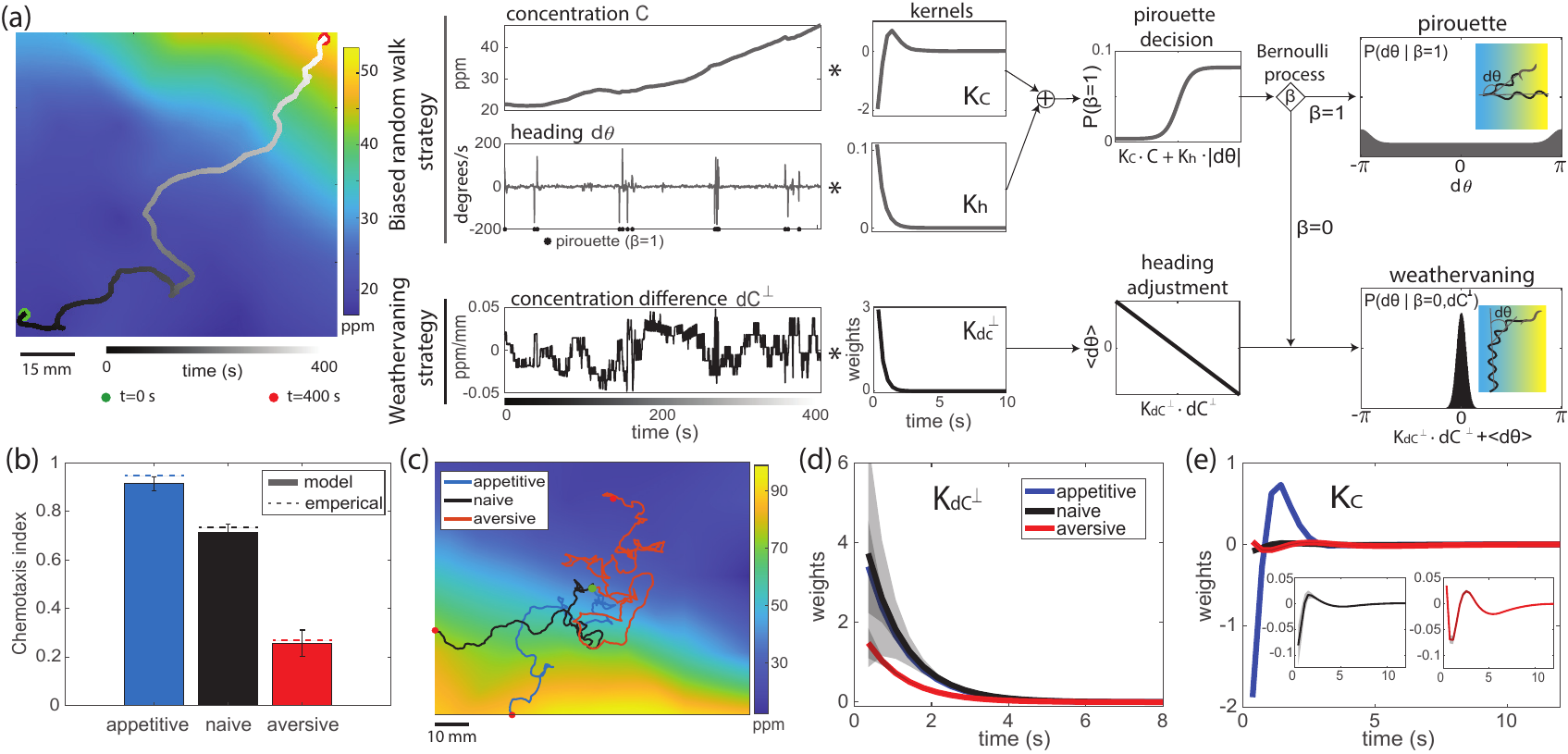}
\caption{dPAW captures learned olfactory navigation in worms. \textbf{(a)} Schematic of the dPAW model. An example data trajectory is shown on the far left, providing time series of concentration and angle changes. Response kernels and decision functions are fitted to the time series data. The Bernoulli process $\beta$ leads to two parallel strategies: biased random walk and weathervaning. \textbf{(b)} Chemotaxis index of simulated trajectories with inferred parameters. Error bar shows standard error of mean across 10 repeated simulation, each with 100 trajectories. \textbf{(c)} Example trajectories simulate from each training conditions. \textbf{(d)} Kernel  $K_{dC^{\perp}}$ and \textbf{(e)} Kernel $K_C$ fitted to three training conditions. Shaded area shows standard deviation of the kernel estimate.}
\label{fig:fig2}
\end{figure*}

To characterize how olfactory sensory inputs are transformed to behavior under different training conditions, we developed a dynamic Pirouette and Weathervaning (dPAW) statistical model of worm olfactory navigation. The dPAW model consists of a mixture of two navigation strategies: a pirouette behavior consisting of an abrupt change in heading angle (a turn) and weathervaning which instead continuously modulates heading angle. 
The dPAW model describes how the worm implements and balances these two behavioral strategies depending on time-varying sensory inputs (Figure \ref{fig:fig2}a). 
The dPAW model is an extension to a classic biased random walk, where the run intervals are replaced with weathervaning behavior.
This framework explicitly models these strategies and is fit to detailed measurements of movement and odor-experience. This is to our knowledge the first  statistical model that explicitly captures the detailed changes to \textit{C. elegans} navigation strategies upon butanone learning.

The model worm samples its heading change $d\theta$ at each time step from one of two distributions: either a ``weathervaning distribution'' $P_{\mathrm{wv}}(d\theta)$ or a ``pirouette distribution'' $P_{\mathrm{pr}}(d\theta)$.
The weathervaning distribution is narrow, reflecting small changes in heading angle that result from the worm's recent measurements of the concentration gradient.
Pirouette behavior, on the other hand, corresponds to large turns that the worm makes when it receives evidence that it is going in the wrong direction. The pirouette distribution is therefore broad, with a peak at $\pm \pi$, indicating a complete reversal in direction. The concentration-dependent ``decision'' to produce a pirouette in between runs forms a biased random walk \cite{Pierce-Shimomura1999-nt, berg1993random}.

The worm's decision to initiate a pirouette or to continue to weathervane on each time step is modeled with a Bernoulli generalized linear model (GLM) that takes the filtered history of the odor concentration $C_{1:t-1}$ and the worm's own movement history $d\theta_{1:t-1}$ as inputs. The output of this GLM is a binary variable $\beta_t$ that indicates the presence of a pirouette. Thus, the worm samples its heading change from the pirouette distribution $P_{\mathrm{pr}}(d\theta)$ if $\beta_t=1$ and the weathervaning distribution $P_{\mathrm{wv}}(d\theta)$ if $\beta_t=0$.
The full model can be written: 
\begin{multline} \label{eq:1}
    P(d\theta_{t} | C_{1:t-1}, dC_{1:t-1}^{\perp}, d\theta_{1:t-1}) = \\
    P(\beta_t=1)P_{\mathrm{pr}}(d\theta)
    + P(\beta_t=0)P_{\mathrm{wv}}(d\theta \mid dC^{\perp}_{1:t-1}), 
\end{multline}
where 
\begin{equation}\label{eq:2}
    P(\beta_t=1) = m+ \frac{M-m}{1+\exp(K_C \cdot \mathbf{C}_{1:t-1} + K_h \cdot |\mathbf{d\theta}_{1:t-1}|)} 
\end{equation}
is the mixing probability over the two distributions, with parameters $m$ and $M$ for the minimum and maximum probability of a pirouette on a single time bin, and $K_C$ and $K_h$ corresponding to filters on past odor concentration $\mathbf{C}_{1:t-1}$ and the past absolute angular change $|\mathbf{d\theta}_{1:t-1}|$ vectors, respectively.  
The pirouette and weathervaning distributions are in turn given by
\begin{align}\label{eq:3}
P_{\mathrm{pr}}(d\theta) &= \alpha U[-\pi, \pi] + (1-\alpha) f(\pi, \kappa_{\mathrm{pr}}), \\
P_{\mathrm{wv}}(d\theta \mid dC^{\perp}_{1:t-1})&= f(-K_{dC^{\perp}} \cdot \mathbf{dC}_{1:t-1}^{\perp}, \kappa_{\mathrm{wv}})
\end{align}
where $U$ is uniform in the circular heading and $f$ is a von Mises distribution with mean and precision parameter $\kappa$. In the pirouette distribution, scalar $\alpha \in [0,1]$ is the weight on the uniform distribution and $\kappa_{\mathrm{pr}}$ is the precision parameter that determines the sharpness of the pirouette. In the weathervaning distribution, the mean is altered according to the perpendicular concentration change $dC^{\perp}_{1:t-1}$ and the precision parameter $\kappa_{\mathrm{wv}}$ determines the noise around the head angle. (Note we have made a simplification by allowing the model access to the instantaneous perpendicular odor concentration. In reality, the animal is thought to compute $dC^{\perp}_{1:t-1}$  from sequential measurements in time as it swings its head through space.)

We fit dPAW to measurements of animal movement in the odor arena, including the time varying headings $d\theta_t$, the concentration along the locomotion path $C_t$, and the concentration perpendicular to the locomotion path $dC^{\perp}_t$. All model parameters in dPAW are jointly inferred through a maximum-likelihood method. To validate that parameters are reliably inferred, we simulated example chemotaxis trajectories from pre-defined parameters, fit them by the model, and confirmed that the model accurately recovered the pre-defined parameters (Fig. S3). 
In the rest of the paper we fit the model to measurements to explore how navigation strategies change with learning.

\subsection*{Model captures navigation  altered by olfactory learning}
To characterize how olfactory learning alters navigation strategies, we fit the dPAW model to trajectories measured in the butanone odor environment after different training conditions. We first confirmed that the model capture key aspects of the animal's navigation. We confirmed that the fitted model's estimate of pirouette frequency matches that measured empirically \cite{Pierce-Shimomura1999-nt} (Figure S4b). 
We also used the model to simulate chemotaxis behavior in the odor environment and confirmed that 
model-generated  trajectories have a chemotaxis index that agrees with measurement and is similarly bidirectionally modulated by training conditions (Figure \ref{fig:fig2}b). Model-generated trajectories also appear visually similar to experimental observations (Figure \ref{fig:fig2}c). And in further agreement, simulated trajectories recapitulate sensory and behavioral statistics measured in experiments, including distributions of the worms: heading angle,  pirouette rate, experienced  perpendicular odor concentration difference experienced, and tangential odor concentration (Fig. S4,5).

Agreement between model and measurement was not due to chance. The dPAW model on average captures $0.3-0.6$ bits/s more information, depending on the animal's training condition, about navigation behavior than a null model that lacks any olfactory sensing mechanisms (Figure \ref{fig:fig3}c). For comparison to additional null models see Fig. S5a.

\subsection*{Sensorimotor kernels change upon learning}
\begin{figure}[t]
\centering
\includegraphics[width=1.\linewidth]{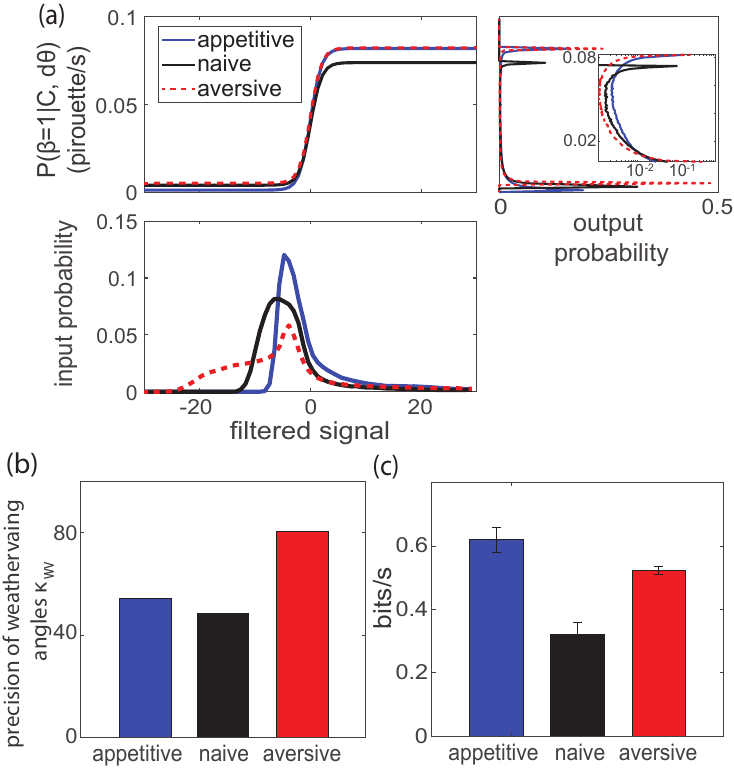}
\caption{Learning alters pirouette decision and behavioral noise. \textbf{(a)} The top panel shows decision function $P(\beta=1|C,d\theta)$ of three training conditions as a function of filtered signal: $K_C\cdot C + K_h \cdot |d\theta|$. The distribution of the filtered signal is shown in the bottom panel. The right panel shows the distribution of the output pirouette rate, with inset showing the same distribution in log scale.
\textbf{(b)} The precision parameter for weathervaning, $\kappa_{\mathrm{wv}}$ across three training conditions. \textbf{(c)} The information rate given fitted model parameters and data across three training conditions. Error bar shows standard error of mean across 10 sampled batches of trajectories.}
\label{fig:fig3}
\end{figure}

It is unknown whether sensorimotor kernels, $K_C$ and $K_{dc^{\perp}}$, should necessarily change upon learning \cite{Pierce-Shimomura1999-nt,Iino2009-al, Cho2016-is}. For example, an animal could in principle change its chemotaxis index upon learning 
by altering its  overall pirouette rate and the curvature of its runs without having to alter the kernels that govern its sensorimotor response. To test this, we inspected the kernels inferred from dPAW that were fit on animals that  underwent different training conditions. Kernels corresponding to both weathervaning and biased random walk strategies are altered by learning (Figure \ref{fig:fig2}d,e). The weathervaning kernel $K_{dC^{\perp}}$ has lower weights after aversive training compared to appetitive trained or naive animals (Figure \ref{fig:fig2}d), suggesting that the animal's heading angle is less tightly dependent upon the concentration difference perpendicular to its path. In other words, aversive trained worms don't pay as much attention to perpendicular concentration when choosing their heading angle. There are two extreme ways these worms could not pay attention to the sensory cue: the worm could be uncoordinated and randomly select a heading direction, or alternatively, it could keep its existing heading. We observe the latter. We found that aversive-trained worms alter the behavioral noise and are more likely to preserve their existing heading and exhibit higher persistent length during their runs, as indicated by higher fitted precision parameter of weathervaning $\kappa_{\mathrm{wv}}$ (Figure \ref{fig:fig3}b).

The kernels corresponding to pirouette decisions, $K_C$, change for both aversive and appetitive training compared to naive animals (Figure \ref{fig:fig2}e). The amplitude is increased after appetitive training compared to naive animals, suggesting that these animal's pirouette probability more strongly depends on the concentration change along the navigation path than for naive animals. After aversive training, the kernel $K_C$ has a longer time delay and forms a tri-phasic shape, markedly different from the biphasic shape observed in naive animals, suggesting that aversive trained animals may not be responding as much  to  downward changes in concentration (Figure \ref{fig:fig2}e).  

Collectively, we show that \textit{C. elegans} alter their chemotaxis upon learning by changing the kernels that govern their sensorimotor response. 


\subsection*{Decision function governing pirouettes changes upon learning}
To understand how the animal alters its preference for continuing to weathervane versus interrupting weathervaning with pirouettes, we compared the pirouette decision function in equation \ref{eq:2} inferred from our measurements before and after learning.

We found that learning alters the input-output statistics governing the initiation of a pirouette. This is clear from inspecting the distribution of the filtered signal (Figure \ref{fig:fig3}a, bottom, related to odor concentration and past behavior) that serves as input to the decision function.  Appetitive training and aversive training pushes the tail of the filtered signal probability distribution in opposite directions with respect to naive condition, which changes the statistics of pirouettes. This could reflect either changes to the kernels, or changes to the environment that those animals prefer to explore.

Aversive-trained worms have higher baseline turning rate $m$ than appetitive trained worms, as seen in the output probability (Figure \ref{fig:fig3}a, right), which is the result of passing the filtered odor signal (Figure \ref{fig:fig3}a, bottom) through a nonlinear decision function (equation \ref{eq:2} and Figure \ref{fig:fig3}a, top).
Both aversive- and appetitive-trained worms have higher maximum pirouette rate $M$ compared to naive worms, reflecting a change to the decision function itself.

It is interesting to note that appetitive (but not aversive) trained animals spend more time with their pirouette probabilities in the most sensitive range of the sigmoid (Fig \ref{fig:fig3}a right, inset), possibly indicating a more efficient strategy for chemotaxis.  Capturing these details is one of the ways that dPAW is able to better detect changes in learning.

\subsection*{Model outperforms other metrics at decoding learned experience}

\begin{figure}[t]
\centering
\includegraphics[width=.8\linewidth]{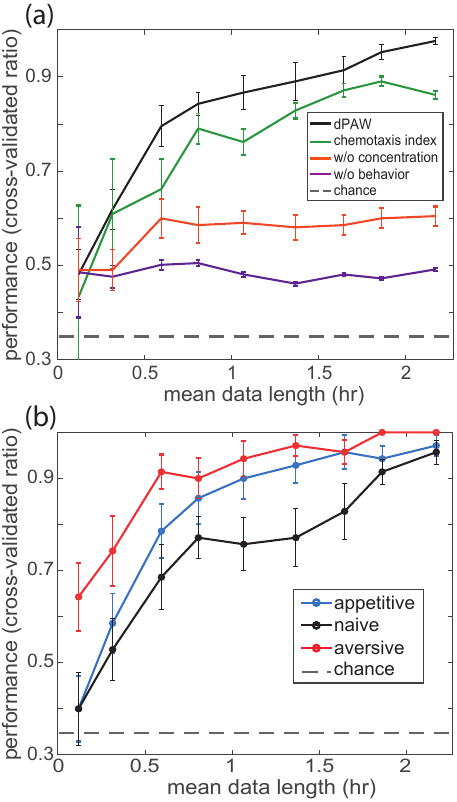}
\caption{Model-based decoding of learning. \textbf{(a)} Model performance classifying the prior training of a population of worms (aversive, appetitive or naive) as a function of mean data length. Four models are compared and error bars show standard error of mean across 7-fold cross validation and 10-fold sampling across data length. w/o concentration: model that only captures behavioral statistics and does not account for odor input. w/o behavior: model for concentration difference along each trajectory and does not account for behavioral output. Chance level, 33\%, is shown in grey dash line. \textbf{(b)} dPAW-based decoding as a function of data length as in (a), but here performance for each training condition is reported separately.}
\label{fig:fig4}
\end{figure}

The changes to sensorimotor processing detected by the dPAW model all provide information about the animal's past experience. We therefore wondered whether the model could accurately predict the training (aversive, appetitive or naive) that a population of worms had experienced. To test this 
we inspected  dPAW models fit to measured trajectories from each training conditions $\gamma \in \{\textrm{appetitive}, \textrm{naive}, \textrm{aversive}\}$  corresponding to fitting parameters $\Theta_\gamma$ and made maximum likelihood predictions of the training condition given held-out test trajectories: $\hat \gamma = \arg \max_\gamma P(\Vec{d\theta}| \Vec{C}, \Vec{dC^{\perp}} ; \Theta_\gamma)$. On held out data, the fitted model correctly predicted training condition with a performance well above 90$\%$,  significantly above chance levels (Figure \ref{fig:fig4}a), given sufficiently long recordings.

The inferred kernels and decision function within $\Theta_\gamma$  better reflect the worm's learning than either a classic chemotaxis index that captures the fraction of trajectories that go up-gradient (Figure \ref{fig:fig4}a), or the concentration difference along the track (Fig. S7). Indeed, dPAW always outperforms the chemotaxis index at decoding past training,  for any tested amount of finite data. The model's predictive power is derived from observing both the odor-history experienced by the animal and the corresponding behavior responses. Models supplied with either odor-history or behavior but not both fail to perform as well.

Some training conditions are more challenging for dPAW to decode then others. Naive worms show the lowest predictive performance with finite data when decoding is performed for each training condition separately (Figure \ref{fig:fig4}b). This is consistent with our estimate of lower information exhibited by the trajectories of naive worms than trained worms (Figure \ref{fig:fig3}c). Practically, this means more measurements are required to capture navigation behavior in naive worms than in trained worms.

\subsection*{The same computations govern natural and optogenetic odor stimuli}
We wondered whether the same underlying computations that governed the animal's response to natural odor stimuli also govern its response to optogenetic-induced sensory stimuli. Optogenetics stimuli are not bound by the same natural statistics that the worm encounters when exploring a physical odor arena. For example, worms in our arena always experience temporally correlated and slowly varying odor responses. This introduces potentially confounding temporal correlations that can be avoided by using optogenetic stimuli \cite{Cho2016-is, Gepner2015-iq, hernandez2015reverse}.
We therefore investigated  response to optogenetic induced odor sensation  after learning (Supplementary Table S2) and adapted our model to incorporate both forms of stimuli.

\begin{figure*}[t]
\centering
\includegraphics[width=.9\linewidth]{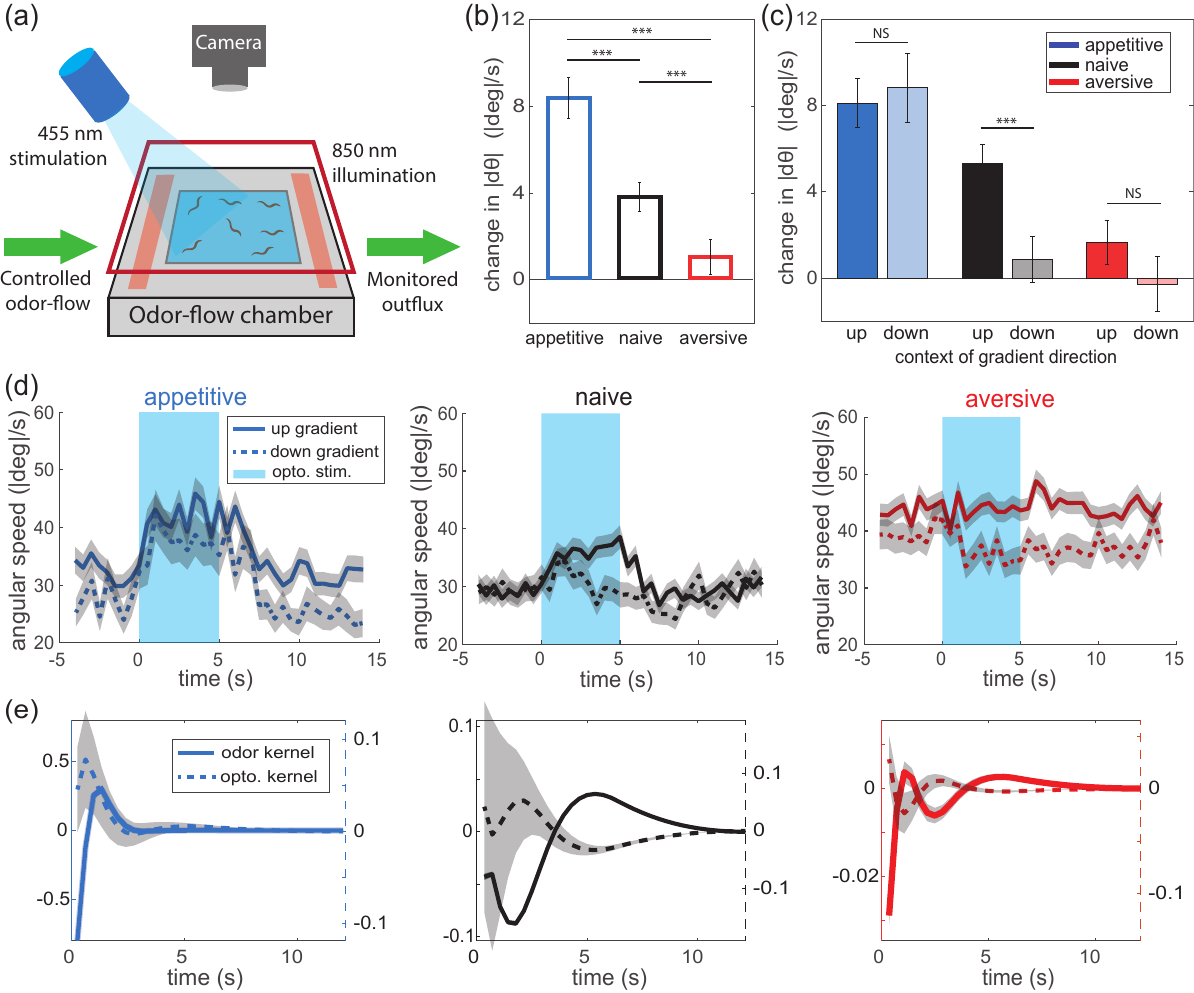}
\caption{Response to optogenetic induced odor sensation is altered by learning and depends on odor gradient context.  \textbf{(a)} Optogenetic stimulation is delivered to animals expressing ChR2 in AWC$^\mathrm{ON}$ as they crawl in an  odor-flow chamber. \textbf{(b)} Change in the absolute heading $|d\theta|$ upon optogenetic impulse across the three training conditions. Change is computed between the average response during the 5s impulse and 5s before it. We record from 4-7 plates per condition. Error bar shows standard error of mean across over 1000 impulses per condition.  The results of t-test comparison between training conditions are shown with $***$ for $p < 0.001$. \textbf{(c)} Same measurements as (b) upon optogenetic impulse, but with the context of gradient direction. Within each training conditions, measurements are separated into trajectories going up or down gradient. The results of t-test comparison between context are shown with $***$ for $p < 0.001$ and NS for non-significance.
\textbf{(d)} Time-varying average angular speed aligned to the optogenetic stimuli. Three panels show different training conditions. Solid line and dash line indicate up and sown gradient trajectories. Gray shading shows standard error of mean across traces. \textbf{(e)} Temporal kernels for odor $K_C$ and optogenetic $K_{opto}$ fitted to each training conditions. The grey area shows standard deviation around the estimated kernels.}
\label{fig:fig5}
\end{figure*}

The animal's behavioral response to optogenetic stimulation was bidirectionally altered by learning (Figure \ref{fig:fig5}, Fig. S8, Fig. S9). We measured the animal's absolute change in heading angle $|d\theta|$ in response to optogenetic stimulation of neuron AWC$^\mathrm{ON}$ expressing ChR2 (Figure \ref{fig:fig5}b). AWC$^\mathrm{ON}$ is a butanone sensitive neuron known to play an important role in learning \cite{Cho2016-is, Pritz2023-du, Kauffman2011-up}. We measured behavior response to optogenetic stimuli two ways: we delivered pulses of optogenetic stimuli to animals experiencing odor in the arena (Figure \ref{fig:fig5}, Fig. S8), and we delivered time-varying intensity white noise optogenetic stimuli to animals off odor (Fig. S9). In both cases (Figure \ref{fig:fig5}e and Fig. S9) the animal's turning behavior was more tightly coupled to optogenetic stimuli for appetitive trained animals than naive worms, and less so for aversive trained animals. 
A change in behavior response is consistent with prior reports \cite{Cho2016-is}, but note that here we make a more stringent comparison by comparing both appetitive and aversive trained animals against the same naive control condition.

We extended our model so that light intensity contributes to the pirouette probability during navigation: 

\begin{equation} \label{eq:5}
    P(\beta_t=1) = \frac{M-m}{1+\exp(K_{odor} \cdot\mathbf{C}_{1:t-1} + K_{opto} \cdot \mathbf{I}_{1:t-1})} + m
\end{equation}
where kernels $K_{odor}$ and $K_{opto}$ are weights on vectors of odor concentration $\mathbf{C}_{1:t-1}$ and light intensity $\mathbf{I}_{1:t-1}$, respectively. Since there is no difference along the perpendicular direction for this five second long spatially uniform optogenetic pulse, this model is simplified with kernels weighting the tangential concentration and neglecting the perpendicular concentration for weathervaning. 

Across different training conditions, the optical kernel is close to the mirror image of the odor kernel, most strikingly demonstrated in the naive and aversive conditions (Figure \ref{fig:fig5}e).  Therefore the sensorimotor computation inferred from freely moving animals is predictive of the response to external perturbations. The inversion is expected and can partly be explained by the biophysics of the AWC$^\mathrm{ON}$ neuron: it hyperpolarizes when odor is present and depolarizes when odor is removed. This gives us confidence that our findings about sensorimotor processing derived from natural odor stimuli should be relevant to the larger literature based on more artificial stimulation. We therefore will leverage optogenetics to probe behavioral variability during odor navigation.

\subsection*{Learning modulates behavioral variability in response to sensory perturbation}
We sought to characterize the variability in learned odor-guided navigation, because variability is a known feature of sensory processing in worms \cite{gordus2015feedback}. We observe variability across collections of trajectories in  learned navigation. For example, the kernel $K_{C}$ that governs the timing of pirouettes varies across subsamples of our data, and seems to vary more for aversive-trained than appetitive-trained or naive animals (Fig. S6).  

Surprisingly, we discovered that animals respond to stimuli differently depending on whether they are traveling up or down an odor gradient and that this contributes to variability (Figure \ref{fig:fig5}c,d). 
Naive worms respond more strongly to optogenetic impulses delivered when traveling up an odor gradient, than when traveling down the gradient  (Figure \ref{fig:fig5}c).
Appetitive trained worms, by contrast, respond consistently to optogenetic impulses regardless of their direction of travel along the odor  gradient. Aversive trained worms show weak response to optogenetic impulse regardless of the gradient context. This indicates that learning alters behavioral variability, and that a response to stimuli can be context-dependent along the navigation trajectory.

\subsection*{Downstream interneurons differentially contribute to learned chemotaxis}
\begin{figure*}
\centering
\includegraphics[width=.8\linewidth]{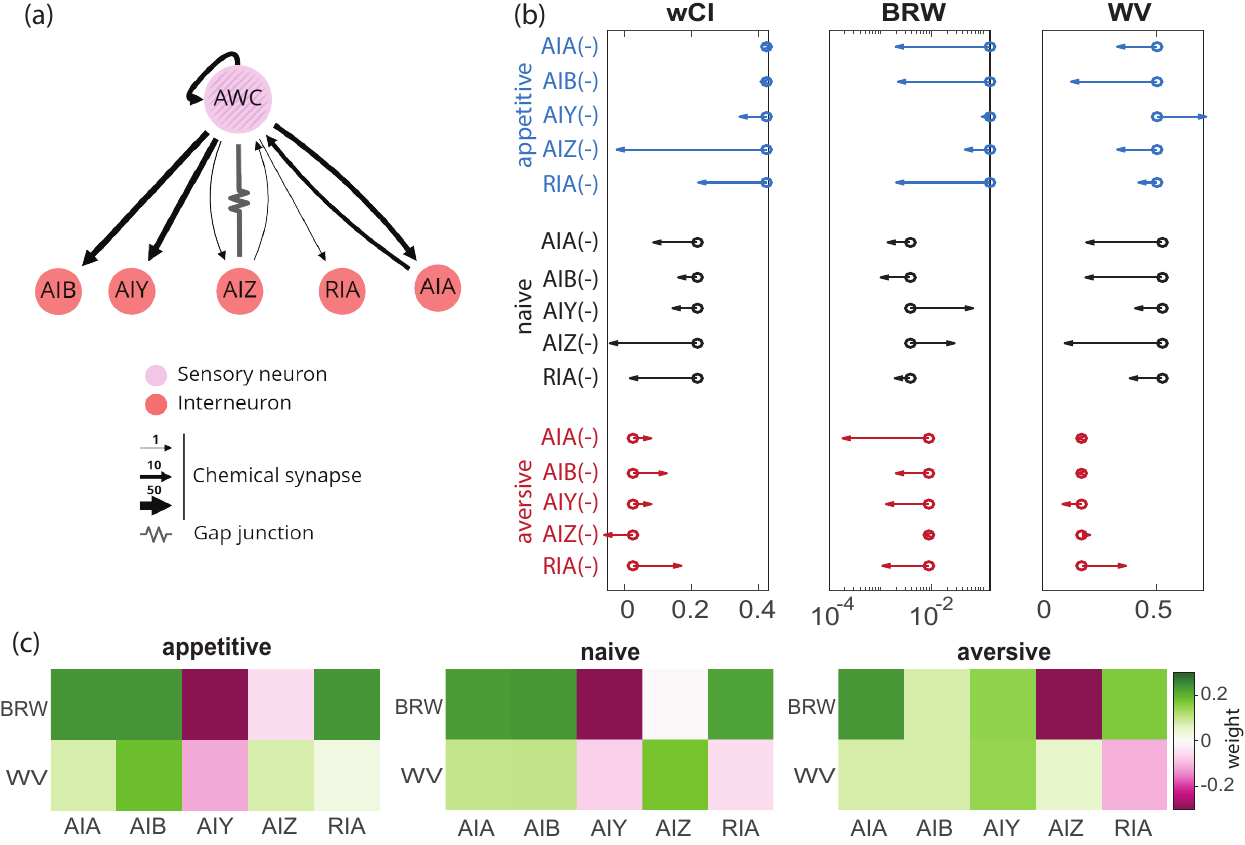}
\caption{Learned odor navigation in worms with disrupted interneurons. \textbf
{(a)} Five interneurons connected to the AWC sensory neuron. Schematic generated from \url{nemanode.org}. \textbf{(b)} Weighted chemotaxis index (wCI), biased random walk (BRW), and weathervaning (WV) indices computed from the inferred dPAW for all mutants across three training conditions. BRW and WV correspond to the norm of kernels $K_C$ and $K_{dC^{\perp}}$ in dPAW. The circles show value from wild-type N2 worms and arrow shows the modulation in mutant worms. We record 3-5 plates for each stain and condition, resulting in 200-500 tracks in each measurement. \textbf{(c)} Weights in the neuron-behavior matrix $\mathbf{W}$ across three training conditions. The two rows are for different behavioral strategies and the columns are for different interneurons. 
}
\label{fig:fig6}
\end{figure*}

We investigated interneurons that may be involved in implementing learned changes to navigation strategy. We focused on interneurons downstream of the odor-sensing neuron AWC$^\mathrm{ON}$ because optogenetically induced activity in AWC$^\mathrm{ON}$ is sufficient to recapitulate aspects of learned navigation (Figure \ref{fig:fig5}b) and AWC$^\mathrm{ON}$ is known to play an important role in odor sensing \cite{Bargmann2006-dy}, navigation \cite{Kato2014-ny, Chalasani2007-oy, Levy2020-oh}, and butanone learning \cite{Torayama2007-qi, Cho2016-is}. We selected five interneurons subtypes with direct synaptic inputs (chemical or electrical connections) from AWC$^\mathrm{ON}$ \cite{Cook2019-dl, white1986structure}: AIA, AIB, AIZ, AIY, and RIA, several of which are known to be involved in navigation \cite{Chalasani2007-oy, Iino2009-al, Hendricks2012-bw, Cho2016-is, Ikeda2020-tw, Luo2014-pc}. For each neuron subtype we compared learned odor-guided navigation in wild type animals to that of  mutants for which the neuron subtype was genetically ablated (via miniSOG, Supplementary Table S1) or down-regulated (via expressing activated potassium channel), as described in methods, Figure \ref{fig:fig6}.

Ablating or down-regulating the downstream interneurons had wide-ranging and statistically significant effects on chemotaxis performance and inferred model parameter, including after learning (Figure \ref{fig:fig6}b; Fig. S10, Fig. S11). Here we characterize chemotaxis performance with a weighted chemotaxis index (wCI) that differs from the chemotaxis index in Figure \ref{fig:fig1}d by more heavily weighting trajectories that experience a large change in odor concentration and de-emphasizing tracks that experience little odor change (described in the methods). To capture changes in inferred model parameters we report a stimuli-normalized magnitude of the kernels $K_C$ and $K_{dC^{\perp}}$ which correspond to the extent to which the animal uses the biased random walk (BRW) or weathervaning strategies (WV), respectively, Figure \ref{fig:fig6}b.

Ablation of neuron subtype AIZ was most severe and eliminated gradient climbing behavior across all three conditions (wCI close to zero), consistent with its reported role in salt chemotaxis \cite{Iino2009-al}.  In general, though, neuron subtypes contributed differently to chemotaxis performance, learning, or had different contributions to the inferred kernels corresponding to each navigation strategies. For instance, AIA, AIB and AIY defective animals showed little difference between naive and aversive training conditions, but did increase chemotaxis performance after appetitive training. The RIA defective animals have similar chemotaxis performance in upon appetitive and aversive learning, suggesting that RIA is be involved in fine-tuning heading angle after learning experience. Interestingly, RIA defect results in much lower gradient climbing performance for naive animals. This suggests that RIA may be related to both learning and sensory adaptation in odor navigation.

We systematically quantified each neuron's contribution to the changes in the inferred behavioral strategies with a linear model  (Figure \ref{fig:fig6}c): $\mathbf{B}=\mathbf{W}\mathbf{A}$, where $\mathbf{B}$ is the matrix with behavior readout (BRW and WV indices) along the rows and ablation conditions along the columns, $\mathbf{W}$ is an unknown weight matrix from neuron to behavior, and $\mathbf{A}$ is the neural ablation matrix with binary element with ones on the off diagonal elements. Matrix $\mathbf{A}$ is full-rank and makes the linear model solvable. We normalized the rows in $\mathbf{B}$ to better compare weights across behaviors.

The majority of the weights $\mathbf{W}$ were positive  (Figure \ref{fig:fig6}c), indicating that a neuron contributes to a given navigation strategy for a given training and that, conversely, disrupting the neuron down-regulates the navigation strategy. Interestingly some neurons, such as RIA, had negative weights. Unlike wild-type animals which decrease their weathervaning upon aversive training, RIA-defective animals fail to adjust the extent of their weathervaning upon aversive training.  This suggests that RIA may act in a learning-dependent manner to decrease or increase odor-dependence of head angle, consistent with RIA's known role in controlling head motion \cite{Hendricks2012-bw}.

The patterns of neuron weights were similar for naive and appetitive conditions, which both corresponded to wild-type positive chemotaxis. The patterns of neuron weights for aversive behavior, however was quite distinct, suggesting that the neural mechanisms of  aversive learning may be quite distinct from that of appetitive learning. 

Interestingly, no neurons had non-zero weights exclusively  on one strategy (e.g. biased random walk) but not the either (e.g. weathervaning),  suggesting that even though these two strategies are mathematically and behaviorally distinct, they do not segregate cleanly into different sub-populations of neurons. Instead the neural basis for learning dependant weathervaning and biased random walks are both distributed across the neural population.

\section*{Discussion}

We combined precise experimental measurements and a novel statistical model, dPAW, to characterize learned olfactory navigation in worms. The results show that (1) navigation strategies are bidirectionally altered by learning, (2) the dPAW model decodes the animal's past training conditions based on the observed navigation behavior and outperforms a classic chemotaxis metric, (3) behavior is more variable in naive worms and this variability can in part be explained by a context-dependent response to odor, and (4) interneurons downstream from AWC$^\mathrm{ON}$ contribute to learned navigation strategies in a distributed manner.  

To reach these conclusions, two experimental advances were critical. The first was the measurement of navigation in a precisely defined sensory environment enabled by our recently developed odor delivery system \cite{Chen2023-fy}. Previous studies had analyzed navigation based on the proximity to a droplet source \cite{Cho2016-is, Torayama2007-qi, Pritz2023-du}, but in those experiments the precise odor concentration experienced by the animal was typically not known. With the odor delivery system, we obtained the odor concentration experienced by the animal and therefore were able to fit more comprehensive models of sensorimotor processing. 

The second methodological advance is the development of a more robust training protocol to probe bivalent learning (aversive and appetitive) that shows clearer effects when tested in the same odor concentration range. Prior investigations into butanone learning probed different valence learning  (appetitive vs. naive, or aversive vs. naive) in different concentration regimes \cite{Kauffman2011-up, Cho2016-is, Torayama2007-qi, Chandra2023-au}, possibly because the behavior effects of aversive learning are known to be more pronounced at lower odor concentrations of butanone while appetitive is known to be more pronounced at higher concentrations of butanone (see for example Fig 1- figure supplement 2 of \cite{Pritz2023-du} and results in \cite{Yoshida2012-dp}).
But here we sought to directly compare detailed properties of learning, such as kernels, in the same sensory environment. To do so, we changed the training protocol (more repetitions during aversive than appetitive) in order to achieve a greater difference in learning outcomes. 
This produced large bivalent training-dependent changes to chemotaxis that were visible even in the same sensory environment  (Figure \ref{fig:fig1}).

The dPAW model fitted to measurements provides new insight into how the animal responds to time varying sensory stimuli. In particular, we find that learning alters the temporal kernels for odor input. For instance, appetitive training sharpens the tangential concentration kernel (Figure \ref{fig:fig2}). By contrast, classical approaches have  missed this important change because they implicitly have static sensing kernels and \textit{a priori} fix the time window 
 to calculated gradients  
\cite{Iino2009-al, Luo2014-pc, Ikeda2020-tw, Pierce-Shimomura1999-nt}. 

Our measurement and model reveals that learning alters not only the sensing mechanism but also the statistics of  behavioral noise (Figure \ref{fig:fig3}), which is consistent with recent work showing that starvation and neuromodulation can dramatically alter the statistics of  behavior \cite{Thapliyal2023-yr, Omura2012-wk}. Our finding was possible only because dPAW explicitly includes parameters that control the noise level of the behavioral output, which is another advantage over past work which often excludes noise and relies on predetermined parameters \cite{Iino2009-al, Dahlberg2020-ip, Pierce-Shimomura1999-nt}.
Future work is needed to pinpoint the source of behavioral stochasticity, such as noise in the sensory or motor circuits, and its potential functional roles in navigation and exploration \cite{Bartumeus2016-kw, gordus2015feedback}.


A strength of our approach is that it allows us to learn properties of sensorimotor kernels from either artificial optogenetic stimulation or natural odor stimuli. Past work, by contrast, required choosing one approach or another \cite{Gepner2015-iq, Kato2014-ny, Cho2016-is}. Optogenetic stimulation has advantages because it can be finely manipulated to deliver rich informative stimuli, for example white noise stimuli to study sensory encoding \cite{Pillow2008-vg, Sharpee2004-zi}. But there are challenges to connect optogenetic stimuli to natural stimuli experienced during navigation. 
Our work shows directly that information from one approach is compatible to the other. We applied statistical inference directly to navigational trajectories in the presence of odor and qualitatively recover similar sensorimotor transformations to those we inferred using optogenetic perturbation (Figure \ref{fig:fig5}). This is to our knowledge the first direct comparison between sensorimotor computation during navigation and response upon external perturbation.

An important conclusion from optogenetic perturbations is that the worm's response variability is altered by learning. We found that naive worms are more variable and that learning reduces variability in behavioral strategies across the population of worms. Surprisingly, we further found that the variation across optogenetic responses in naive worms can in part be explained by the context of that worm's odor environment, up or down the gradient (Figure \ref{fig:fig5}). Naive worms that travel up gradient pay more attention to both odor and optogenetic stimuli (Figure \ref{fig:fig5}c, Fig. S8) compared to those that go down gradient. 
This is broadly consistent with prior literature describing how the result of associative learning can be heterogeneous across animals or non-stationary in time \cite{Lesar2021-lo, Tait2019-sb, Smith2022-gb}.  
Measurements in response to optogenetically perturbing AWC$^\mathrm{ON}$ suggest that the source of such context-dependent behavior might be along the AWC$^\mathrm{ON}$ pathway. Future work is needed to address the source of such variability in the nervous system, as well as its possible functional role for navigation and searching behavior \cite{gordus2015feedback, baker2018algorithms}.

Of the interneurons we investigated, all five classes contribute to both learned navigation strategies (Figure \ref{fig:fig6}). Importantly, many connections from AWC$^\mathrm{ON}$ to the first layer of interneurons are shared with salt sensing neurons ASER/L and thermal sensing neuron AFD \cite{Iino2009-al, Ikeda2020-tw, Luo2014-pc}. Some past work hypothesize that experience-dependent changes are localized in specific neurons---- AIB has been shown to play a role in context-dependent thermal navigation \cite{Ikeda2020-tw}. But other work presents evidence that learning effects can be distributed across many neurons \cite{Luo2014-pc, Pritz2023-du}. For instance, AIB, AIY, and AIZ all seem to be involved in learned salt chemotaxis strategies \cite{Luo2014-pc}. Our findings agree with these results in showing that all three interneurons have non-zero and learning-dependent weights for both behavioral strategies. 


One puzzling finding is that we sometimes notice changed behavioral strategies but unchanged chemotaxis index. For example, appetitive trained worms with defects in AIB have dramatic reductions in both weathervaning and biased random walk strategies but still show reasonable chemotaxis performance. This may be due to larger model mismatch in mutant worms with strategies deviating from dPAW. Large variability and less consistent behavioral strategies were also observed in earlier ablation studies \cite{Iino2009-al}. We explore this further in the supplement (SI Limitations and future work).

In this work we utilize a controlled odor environment and an innovative model to characterize learned odor navigation in worms. The combined approach of precisely delivered sensory stimuli, behavior quantification and navigational modeling continues to be a powerful approach \cite{clark2013mapping, calhoun2017quantifying, datta2019computational} that can be generalized to other sensory modalities and species to study adaptive sensory navigation. By identifying the specific features of behavior that are altered by learning, our investigation lays the groundwork for followup neural imaging studies and will guide our search for neural representations of learning \cite{nguyen2016whole, Roman2023-ub}.

\matmethods{
\subsection*{Worm strains and preparation}
All worms were maintained at 20 C on nematode growth medium (NGM) agar plates seeded with \textit{E. coli} (OP50). We used N2 bristol as wild type worms. A detailed strain list is provided in Supplementary Information (Table S1). Strain AML105 used for optogenetics experiments was integrated using strain CX14418 from \cite{Cho2016-is} employing UV irradiation and was outcrossed six times before testing. AIB(-) strain is from \cite{Chen2023-fy} and AIA(-), AIY(-), AIZ(-) and RIA(-) strains are from \cite{Ikeda2020-tw}. 

Before chemotaxis experiments, we bleached and centrifuged batches of worms to synchronize the next generation as described in \cite{Liu2018-mv}.
L1 synchronized worms were plated to seeded NGM plates on the next day. Experiments were conducted 3 days after seeding, which corresponds to synchronized 1-day-old adult worms. For optogenetic strains, L1 stage worms were plated onto 9 cm NGM agar plates seeded with 1ml OP50 food with 10 $\mu$l all-trans retinal (from 100 mM stock). For interneuron perturbation, miniSOG strains were treated with square wave pulses of 2.16 mW/mm$^2$ 450 nm blue light at 1 Hz for 30 minutes at L1 stage and then allowed to recover before testing.

\subsection*{Olfactory learning protocol}
The olfactory learning protocol was developed by modifying from previous literature \cite{Cho2016-is, Pritz2023-du, Kauffman2011-up, Torayama2007-qi}.
Synchronized young adult worms were removed from food and washed three times with S. Basal solution \cite{Bargmann2006-dy}. Appetitive trained worms were suspended in 10 ml of S. Basal solution on a shaker to starve for 1 hour. After starvation, worms were placed on a 9 cm NGM agar plate with 1 ml of OP50 and 12 $\mu$l of pure butanone (2-Butanone, +99\%, Extra Pure, Thermo Scientific) dropped on the interior of the lid and sealed with Parafilm. To hold butanone droplets in place, we placed 3 agar plugs on the lid and dropped 4 $\mu$l of butanone onto each plug.
To conduct aversive training, worms were suspended in 10 ml of S. Basal with 1 $\mu$l butanone added. The tube was sealed and placed on the shaker for 1 hour. 
We found that aversive training protocol was most robust with repetition (Fig. S2), so we interleaved the session by plating the worms back on food for 30 minutes, then repeating the training for three times (Figure \ref{fig:fig1}a). Worms were washed three times with S. Basal solution and centrifuged between each transfer during the training protocol. For naive condition, worms were directly removed from food and washed for three times before testing directly. 

\subsection*{Odor delivery system and chemotaxis experiments}
We used the odor flow delivery system and experimental protocols developed in \cite{Chen2023-fy} to measure chemotaxis trajectories after different training conditions. In short, this system incorporates controlled odorized airflow and continuously measures odor concentration along the boundary of an agar plate during animal experiments. As in \cite{Chen2023-fy}, we calibrated the full array of metal-oxide based sensors with a downstream photo-ionization detector (Fig. S1a) to characterize the steady-state spatial profile (Fig. S1b). During animal experiments, we swapped out sensors in the middle of the arena to place in worms on an agar plate and continued to measure the boundary condition to confirm that the odor landscape is controlled and stable across the arena (Fig. S1c), as in \cite{Chen2023-fy} .

For chemotaxis experiments, we used 1.6 $\%$ agar with salt content matching S. Basal solution in a 10 cm square plate lid. We used 11 mM butanone dissolved in water as the odor reservoir. Moisturized clean air as the background flow. The background airflow is 400 ml/min and the butanone odor flow is 33-36 ml/min. After the pre-equilibration protocol \cite{Chen2023-fy} that brings the agar plate to steady-state in the odor environment, 50-100 worms were placed in the middle of agar plate and dried with kimwipes. Each chemotaxis sessions were record for 30 minutes.

\subsection*{Behavioral imaging and optical setup}

Worm behavior imaging was performed as described in  \cite{Chen2023-fy} . Briefly a CMOS camera measured worm behavior at 14 Hz. Worms were illuminated with 850 nm light. Images were captured with custom written Labview program and analyzed with Matlab scripts. An important difference from \cite{Chen2023-fy}, is that here we added the ability to deliver optogenetic stimulation. Three 455 nm LEDs (M455L4, Thorlabs) were fixed on top of the flow chamber to deliver stimulation. The intensity was calibrated in the field of view with a photometer, with 85 $\mu W$/mm$^2$ for each uniform light impulse. Each pulse last for 5 seconds and were delivered every 30 seconds. Here we analyzed 1220-3060 pulses delivered to worms treated with retinal in each training conditions.

\subsection*{Behavioral analysis and dPAW inference}
We tracked the location of the worm's centroid and fit a centerline to its posture as described in \cite{Chen2023-fy}. We removed trajectories that are shorted than 1 minute or have displacement less than 3 mm across the recordings. In addition, trajectories that started above 70$\%$ of the maximum odor concentration were also removed to prevent double-counting worms that may have already started or traveled up-gradient. This results in 270-1,140 animal hours of chemotaxis trajectory per training condition for model fitting.  For each processed navigation trajectory, we computed the displacement vectors every 5 time bins (5/14 seconds) and compute the angle between consecutive vectors to obtain $d\theta_t$. We computed the odor concentration it passes through using the two-dimensional odor landscape measured in the flow chamber $C_t$. The perpendicular concentration difference $dC^{\perp}$ is calculated with unit vectors that are orthogonal to each displacement vector. We also recorded the length of each displacement vector to form the empirical speed distribution.

We fit dPAW to the ensemble of trajectories by maximizing the log-likelihood: 
\begin{equation}
    \mathrm{argmax}  \Sigma_n^N \Sigma_t^T \log P(d\theta_{n,t+1}| C_{n,:t}, dC^{\perp}_{n,:t}, d\theta_{n,:t}) - \lambda |K_C|^2
\end{equation}
where $N$ is the number of trajectories, $T$ is the time steps, and $\lambda$ is the regularization term. To impose smoothness on kernels, $K_C$ is parameterized with 4 raised-cosine basis function \cite{Pillow2008-vg}, $K_{dC^{\perp}}$ and $K_h$ are parameterized with an exponential form. Optimization was performed with constrained optimization in Matlab, where the constrains are positivity of the precision parameters and sigmoid probabilities. Uncertainty about the inferred parameters are characterized by numerically computing the Hessian of the log-likelihood function around the maximum likelihood estimation. For model-based decoding, we perform 7-fold cross validation with all measured trajectories. To test with finite data, we subsample 10 ensembles of trajectories to test for performance in Figure \ref{fig:fig4}.

To validate the accuracy of maximum likelihood inference for dPAW model parameters, we simulated behavioral time series from dPAW with a fixed set of ground-truth parameters. The model generated simulated trajectories using Gaussian random white noise for concentration time series $C_t$ and $dC^{\perp}_t$. The time series has 50,000 data points, which is in the same scale of our data length ($\sim$ 300 animal hours of recording at $5/14$ Hz sampling frequency). We confirmed that the inference procedure works with simulated data and recovers the ground truth parameters (Fig. S3).

To generate chemotaxis trajectories from inferred parameters shown in Figure \ref{fig:fig2}, we conduct agent-based simulation by measuring concentration in the same odor landscape and drawing angular change $d\theta_t$ from dPAW. We simulate two-dimensional navigation trajectories with:
\begin{align}
    x_{t+1} &= x_t + v_t \cos (\theta_t) \\
    y_{t+1} &= y_t + v_t \sin (\theta_t) \\
    \theta_{t+1} &= \theta_t + d\theta_t
\end{align}
where $v_t$ is speed drawn from Gaussian fit to the empirical distribution.

For information rate (Figure \ref{fig:fig3}) and model comparison (Figure \ref{fig:fig4}), we construct null models to compare with dPAW. The random walk model has similar statistical structure for behavior but is independent to odor concentration:
\begin{equation}
    P(d\theta_t) = P(\beta=1)P_{\mathrm{pr}}(d\theta) + P(\beta=0)P_{run}(d\theta)
\end{equation}
Note that in this null model, the turning probability is time-independent, so $\beta$ does not have a time subscript $t$. The pirouette behavior has the similar sharp angle as the full model, but now in the null model the weathervaning is changed to ``runs'' that have zero-mean and do not take perpendicular concentration change into account. This model was fitted to the same ensemble of trajectories, and the log-likelihood difference between the full dPAW and this null model is normalized by $\log(2)$ per time to compute the bit rate. We also used this model to compute behavior-only model prediction in figure \ref{fig:fig4}. The chemotaxis model is formulated with a binomial distribution with expected fraction of track going up gradient $\hat{p}$. The estimation for chemotaxis index is then $2\hat{p} - 1$. Lastly, the concentration change model takes $C_{final}-C_{initial}$ for all tracks and uses naive Bayes classifier for prediction.

\subsection*{Statistical analysis for chemotaxis in mutant worms}
We computed the concentration weighted chemotaxis index: $(N_{up}\Delta C_{up} - N_{down}\Delta C_{down}) / (N_{up}\Delta C_{up
} + N_{down}\Delta C_{down})$, where $N$ is the number of tracks going up or down gradient and $\Delta C$ is computed for every track. We find the maximum and minimum concentration across the full observed odor landscape, then define $\Delta C_{up} = C_{max} - C_{final}
$ for tracks going up gradient and $\Delta C_{down} = C_{final} - C_{min}$ for tracks going down gradient, with ending concentration $C_{final}$. 

To conduct statistical tests between indices in Figure 6, we re-sampled chemotaxis trajectories in each conditions (Fig. S11). For the weighted chemotaxis index, we sampled 50 tracks for 20 times to compute the standard deviation from mean. For the biased random walk strategy, we computed $|K_C|/std(C_t)$ to quantify the weights on concentration $C$ and normalized by the standard deviation of the time series itself to compare across strains that experience different concentration inputs. For the weathervaning strategy, we computed $\mathrm{std}(K_{dC^{\perp}} * dC^{\perp}_{:t})$ to characterize how much the worm weights and corrects for the heading in response to $dC^{\perp}$. 
To conduct statistical tests on the behavioral strategies, we sampled 100 times from the posteriors of these kernels, with Gaussian approximation around the maximum likelihood estimate shown in Figure 2 b,c. For each transgenic strain and training condition, we computed the standard deviation of the metrics and conducted t-test between the transgenic strain and wild type worms (Fig. S11).

\subsection*{Mapping the behavior-triggered average with optogenetics}
For behavior triggered averages (Fig. S9) we delivered time varying LED light stimuli drawn from $\mathcal{N}(30,30)$ ($\mu$W/mm$^2$) at 14Hz with 0.5s correlation time and bounded between 0-60 $\mu$W/mm$^2$. This serves as a white noise stimulus. We computed the behavior triggered average (BTA) for reversals following methods applied to touch sensation in worms \cite{Liu2018-mv}.

}

\showmatmethods{} 

\acknow{We thank the Leifer lab and Pillow lab for helpful discussions. We also like to thank Prof. Cori Bargmann for sharing worm strain CX14418 and Prof. Ikue Mori for sharing worm strains IK2962, IK3240, IK3241 and IK3289. This work was supported by the National Institutes of Health National Institute of Neurological Disorders and Stroke under New Innovator award (DP2-NS116768), the Simons Foundation under award SCGB (543003), an NSF CAREER Award to AML (NSF PHY-1748958), and through the Center for the Physics of Biological Function (PHY-1734030).}

\showacknow{} 

\bibliography{OdorLearn_ref}

\begin{thebibliography}{10}

\bibitem{baker2018algorithms}
KL Baker, et~al., Algorithms for olfactory search across species.
\newblock {\em\protect\JournalTitle{Journal of Neuroscience}} \textbf{38},
  9383--9389 (2018).

\bibitem{porter2007mechanisms}
J Porter, et~al., Mechanisms of scent-tracking in humans.
\newblock {\em\protect\JournalTitle{Nature neuroscience}} \textbf{10}, 27--29
  (2007).

\bibitem{Torayama2007-qi}
I Torayama, T Ishihara, I Katsura, Caenorhabditis elegans integrates the
  signals of butanone and food to enhance chemotaxis to butanone.
\newblock {\em\protect\JournalTitle{J. Neurosci.}} \textbf{27}, 741--750
  (2007).

\bibitem{gire2016mice}
DH Gire, V Kapoor, A Arrighi-Allisan, A Seminara, VN Murthy, Mice develop
  efficient strategies for foraging and navigation using complex natural
  stimuli.
\newblock {\em\protect\JournalTitle{Current Biology}} \textbf{26}, 1261--1273
  (2016).

\bibitem{Zhang2005-if}
Y Zhang, H Lu, CI Bargmann, Pathogenic bacteria induce aversive olfactory
  learning in caenorhabditis elegans.
\newblock {\em\protect\JournalTitle{Nature}} \textbf{438}, 179--184 (2005).

\bibitem{Ha2010-ja}
HI Ha, et~al., Functional organization of a neural network for aversive
  olfactory learning in caenorhabditis elegans.
\newblock {\em\protect\JournalTitle{Neuron}} \textbf{68}, 1173--1186 (2010).

\bibitem{krakauer2017neuroscience}
JW Krakauer, AA Ghazanfar, A Gomez-Marin, MA MacIver, D Poeppel, Neuroscience
  needs behavior: correcting a reductionist bias.
\newblock {\em\protect\JournalTitle{Neuron}} \textbf{93}, 480--490 (2017).

\bibitem{meister2022learning}
M Meister, Learning, fast and slow.
\newblock {\em\protect\JournalTitle{Current opinion in neurobiology}}
  \textbf{75}, 102555 (2022).

\bibitem{clark2013mapping}
DA Clark, L Freifeld, TR Clandinin, Mapping and cracking sensorimotor circuits
  in genetic model organisms.
\newblock {\em\protect\JournalTitle{Neuron}} \textbf{78}, 583--595 (2013).

\bibitem{white1986structure}
J White, E Southgate, J Thomson, S Brenner, The structure of the nervous system
  of the nematode caenorhabditis elegans.
\newblock {\em\protect\JournalTitle{Philosophical Transactions of the Royal
  Society of London Series B}} \textbf{314}, 1--340 (1986).

\bibitem{Bargmann2006-dy}
CI Bargmann, Chemosensation in c. elegans.
\newblock {\em\protect\JournalTitle{WormBook}} pp. 1--29 (2006).

\bibitem{Pritz2023-du}
C Pritz, et~al., Principles for coding associative memories in a compact neural
  network.
\newblock {\em\protect\JournalTitle{Elife}} \textbf{12} (2023).

\bibitem{Cook2019-dl}
SJ Cook, et~al., Whole-animal connectomes of both caenorhabditis elegans sexes.
\newblock {\em\protect\JournalTitle{Nature}} \textbf{571}, 63--71 (2019).

\bibitem{dusenbery1975chemotaxis}
DB Dusenbery, RE Sheridan, RL Russell, Chemotaxis-defective mutants of the
  nematode caenorhabditis elegans.
\newblock {\em\protect\JournalTitle{Genetics}} \textbf{80}, 297--309 (1975).

\bibitem{croll1977sensory}
NA Croll, Sensory mechanisms in nematodes.
\newblock {\em\protect\JournalTitle{Annual Review of Phytopathology}}
  \textbf{15}, 75--89 (1977).

\bibitem{Pierce-Shimomura1999-nt}
JT Pierce-Shimomura, TM Morse, SR Lockery, The fundamental role of pirouettes
  in caenorhabditis elegans chemotaxis.
\newblock {\em\protect\JournalTitle{J. Neurosci.}} \textbf{19}, 9557--9569
  (1999).

\bibitem{Kauffman2011-up}
A Kauffman, et~al., C. elegans positive butanone learning, short-term, and
  long-term associative memory assays.
\newblock {\em\protect\JournalTitle{J. Vis. Exp.}} (2011).

\bibitem{Iino2009-al}
Y Iino, K Yoshida, Parallel use of two behavioral mechanisms for chemotaxis in
  caenorhabditis elegans.
\newblock {\em\protect\JournalTitle{J. Neurosci.}} \textbf{29}, 5370--5380
  (2009).

\bibitem{Ikeda2020-tw}
M Ikeda, et~al., Context-dependent operation of neural circuits underlies a
  navigation behavior in caenorhabditis elegans.
\newblock {\em\protect\JournalTitle{Proc. Natl. Acad. Sci. U. S. A.}}
  \textbf{117}, 6178--6188 (2020).

\bibitem{macnab1972gradient}
RM Macnab, DE Koshland~Jr, The gradient-sensing mechanism in bacterial
  chemotaxis.
\newblock {\em\protect\JournalTitle{Proceedings of the National Academy of
  Sciences}} \textbf{69}, 2509--2512 (1972).

\bibitem{keller1971model}
EF Keller, LA Segel, Model for chemotaxis.
\newblock {\em\protect\JournalTitle{Journal of theoretical biology}}
  \textbf{30}, 225--234 (1971).

\bibitem{berg1993random}
HC Berg, {\em Random walks in biology}.
\newblock (Princeton University Press), (1993).

\bibitem{Luo2014-pc}
L Luo, et~al., Dynamic encoding of perception, memory, and movement in a c.
  elegans chemotaxis circuit.
\newblock {\em\protect\JournalTitle{Neuron}} \textbf{82}, 1115--1128 (2014).

\bibitem{Jang2019-np}
MS Jang, Y Toyoshima, M Tomioka, H Kunitomo, Y Iino, Multiple sensory neurons
  mediate starvation-dependent aversive navigation in caenorhabditis elegans.
\newblock {\em\protect\JournalTitle{Proc. Natl. Acad. Sci. U. S. A.}} (2019).

\bibitem{Colbert1995-vh}
HA Colbert, CI Bargmann, Odorant-specific adaptation pathways generate
  olfactory plasticity in c. elegans.
\newblock {\em\protect\JournalTitle{Neuron}} \textbf{14}, 803--812 (1995).

\bibitem{Tanimoto2017-pt}
Y Tanimoto, et~al., Calcium dynamics regulating the timing of decision-making
  in c. elegans.
\newblock {\em\protect\JournalTitle{Elife}} \textbf{6} (2017).

\bibitem{Chen2023-fy}
KS Chen, R Wu, MH Gershow, AM Leifer, Continuous odor profile monitoring to
  study olfactory navigation in small animals.
\newblock {\em\protect\JournalTitle{Elife}} \textbf{12} (2023).

\bibitem{tadres2022depolarization}
D Tadres, PH Wong, T To, J Moehlis, M Louis, Depolarization block in olfactory
  sensory neurons expands the dimensionality of odor encoding.
\newblock {\em\protect\JournalTitle{Science Advances}} \textbf{8}, eade7209
  (2022).

\bibitem{gershow2012controlling}
M Gershow, et~al., Controlling airborne cues to study small animal navigation.
\newblock {\em\protect\JournalTitle{Nature methods}} \textbf{9}, 290--296
  (2012).

\bibitem{Worthy2018-xi}
SE Worthy, et~al., Identification of attractive odorants released by preferred
  bacterial food found in the natural habitats of c. elegans.
\newblock {\em\protect\JournalTitle{PLoS One}} \textbf{13}, e0201158 (2018).

\bibitem{Cho2016-is}
CE Cho, C Brueggemann, ND L'Etoile, CI Bargmann, Parallel encoding of sensory
  history and behavioral preference during caenorhabditis elegans olfactory
  learning.
\newblock {\em\protect\JournalTitle{eLife Sciences}} \textbf{5}, e14000 (2016).

\bibitem{Lin2023-gf}
A Lin, et~al., Functional imaging and quantification of multineuronal olfactory
  responses in c. elegans.
\newblock {\em\protect\JournalTitle{Sci Adv}} \textbf{9}, eade1249 (2023).

\bibitem{Gepner2015-iq}
R Gepner, M Mihovilovic~Skanata, NM Bernat, M Kaplow, M Gershow, Computations
  underlying drosophila photo-taxis, odor-taxis, and multi-sensory integration.
\newblock {\em\protect\JournalTitle{Elife}} \textbf{4} (2015).

\bibitem{hernandez2015reverse}
L Hernandez-Nunez, et~al., Reverse-correlation analysis of navigation dynamics
  in drosophila larva using optogenetics.
\newblock {\em\protect\JournalTitle{Elife}} \textbf{4}, e06225 (2015).

\bibitem{gordus2015feedback}
A Gordus, N Pokala, S Levy, SW Flavell, CI Bargmann, Feedback from network
  states generates variability in a probabilistic olfactory circuit.
\newblock {\em\protect\JournalTitle{Cell}} \textbf{161}, 215--227 (2015).

\bibitem{Kato2014-ny}
S Kato, Y Xu, CE Cho, LF Abbott, CI Bargmann, Temporal responses of c. elegans
  chemosensory neurons are preserved in behavioral dynamics.
\newblock {\em\protect\JournalTitle{Neuron}} \textbf{81}, 616--628 (2014).

\bibitem{Chalasani2007-oy}
SH Chalasani, et~al., Dissecting a circuit for olfactory behaviour in
  caenorhabditis elegans.
\newblock {\em\protect\JournalTitle{Nature}} \textbf{450}, 63--70 (2007).

\bibitem{Levy2020-oh}
S Levy, CI Bargmann, An {Adaptive-Threshold} mechanism for odor sensation and
  animal navigation.
\newblock {\em\protect\JournalTitle{Neuron}} \textbf{105}, 534--548.e13 (2020).

\bibitem{Hendricks2012-bw}
M Hendricks, H Ha, N Maffey, Y Zhang, Compartmentalized calcium dynamics in a
  c. elegans interneuron encode head movement.
\newblock {\em\protect\JournalTitle{Nature}} \textbf{487}, 99--103 (2012).

\bibitem{Chandra2023-au}
R Chandra, et~al., Sleep is required to consolidate odor memory and remodel
  olfactory synapses.
\newblock {\em\protect\JournalTitle{Cell}} \textbf{186}, 2911--2928.e20 (2023).

\bibitem{Yoshida2012-dp}
K Yoshida, et~al., Odour concentration-dependent olfactory preference change in
  c. elegans.
\newblock {\em\protect\JournalTitle{Nat. Commun.}} \textbf{3}, 739 (2012).

\bibitem{Thapliyal2023-yr}
S Thapliyal, I Beets, DA Glauser, Multisite regulation integrates multimodal
  context in sensory circuits to control persistent behavioral states in c.
  elegans.
\newblock {\em\protect\JournalTitle{Nat. Commun.}} \textbf{14}, 3052 (2023).

\bibitem{Omura2012-wk}
DT Omura, DA Clark, ADT Samuel, HR Horvitz, Dopamine signaling is essential for
  precise rates of locomotion by c. elegans.
\newblock {\em\protect\JournalTitle{PLoS One}} \textbf{7}, e38649 (2012).

\bibitem{Dahlberg2020-ip}
BA Dahlberg, EJ Izquierdo, Contributions from parallel strategies for spatial
  orientation in c. elegans.
\newblock {\em\protect\JournalTitle{ALIFE 2020: The 2020 Conference on}}
  (2020).

\bibitem{Bartumeus2016-kw}
F Bartumeus, et~al., Foraging success under uncertainty: search tradeoffs and
  optimal space use.
\newblock {\em\protect\JournalTitle{Ecol. Lett.}} \textbf{19}, 1299--1313
  (2016).

\bibitem{Pillow2008-vg}
JW Pillow, et~al., Spatio-temporal correlations and visual signalling in a
  complete neuronal population.
\newblock {\em\protect\JournalTitle{Nature}} \textbf{454}, 995--999 (2008).

\bibitem{Sharpee2004-zi}
T Sharpee, NC Rust, W Bialek, Analyzing neural responses to natural signals:
  maximally informative dimensions.
\newblock {\em\protect\JournalTitle{Neural Comput.}} \textbf{16}, 223--250
  (2004).

\bibitem{Lesar2021-lo}
A Lesar, J Tahir, J Wolk, M Gershow, Switch-like and persistent memory
  formation in individual drosophila larvae.
\newblock {\em\protect\JournalTitle{Elife}} \textbf{10} (2021).

\bibitem{Tait2019-sb}
C Tait, A Mattise-Lorenzen, A Lark, D Naug, Interindividual variation in
  learning ability in honeybees.
\newblock {\em\protect\JournalTitle{Behav. Processes}} \textbf{167}, 103918
  (2019).

\bibitem{Smith2022-gb}
MAY Smith, KS Honegger, G Turner, B de~Bivort, Idiosyncratic learning
  performance in flies.
\newblock {\em\protect\JournalTitle{Biol. Lett.}} \textbf{18}, 20210424 (2022).

\bibitem{calhoun2017quantifying}
AJ Calhoun, M Murthy, Quantifying behavior to solve sensorimotor
  transformations: advances from worms and flies.
\newblock {\em\protect\JournalTitle{Current opinion in neurobiology}}
  \textbf{46}, 90--98 (2017).

\bibitem{datta2019computational}
SR Datta, DJ Anderson, K Branson, P Perona, A Leifer, Computational
  neuroethology: a call to action.
\newblock {\em\protect\JournalTitle{Neuron}} \textbf{104}, 11--24 (2019).

\bibitem{nguyen2016whole}
JP Nguyen, et~al., Whole-brain calcium imaging with cellular resolution in
  freely behaving caenorhabditis elegans.
\newblock {\em\protect\JournalTitle{Proceedings of the National Academy of
  Sciences}} \textbf{113}, E1074--E1081 (2016).

\bibitem{Roman2023-ub}
A Roman, K Palanski, I Nemenman, WS Ryu, A dynamical model of c. elegans
  thermal preference reveals independent excitatory and inhibitory learning
  pathways.
\newblock {\em\protect\JournalTitle{Proc. Natl. Acad. Sci. U. S. A.}}
  \textbf{120}, e2215191120 (2023).

\bibitem{Liu2018-mv}
M Liu, AK Sharma, JW Shaevitz, AM Leifer, Temporal processing and context
  dependency in caenorhabditis elegans response to mechanosensation.
\newblock {\em\protect\JournalTitle{Elife}} \textbf{7} (2018).

\end{thebibliography}


\begin{thebibliography}{1}

\bibitem{Ikeda2020-tw}
M Ikeda, et~al., Context-dependent operation of neural circuits underlies a
  navigation behavior in caenorhabditis elegans.
\newblock {\em\protect\JournalTitle{Proc. Natl. Acad. Sci. U. S. A.}}
  \textbf{117}, 6178--6188 (2020).

\bibitem{Pierce-Shimomura1999-nt}
JT Pierce-Shimomura, TM Morse, SR Lockery, The fundamental role of pirouettes
  in caenorhabditis elegans chemotaxis.
\newblock {\em\protect\JournalTitle{J. Neurosci.}} \textbf{19}, 9557--9569
  (1999).

\bibitem{Iino2009-al}
Y Iino, K Yoshida, Parallel use of two behavioral mechanisms for chemotaxis in
  caenorhabditis elegans.
\newblock {\em\protect\JournalTitle{J. Neurosci.}} \textbf{29}, 5370--5380
  (2009).

\bibitem{Chen2023-fy}
KS Chen, R Wu, MH Gershow, AM Leifer, Continuous odor profile monitoring to
  study olfactory navigation in small animals.
\newblock {\em\protect\JournalTitle{Elife}} \textbf{12} (2023).

\end{thebibliography}

\end{document}



\maketitle 

\SItext

\section{Limitations and future work}

With the dynamic Pirouette and Weathervaning (dPAW) model fitted to experimental data, we confirmed that the inferred densities for head angles $d\theta$ matches the observed distribution (Fig. \ref{fig:SI4}a). Two mixtures of von Mises distributions in dPAW can approximate the features observed from experiments. We find the agreement to be overally very good.  But we noticed that where there is mismatch, it is largest in a range of intermediate angles. It is possible that this mismatch could be explained by more subtle aspects of locomotion that have previously been proposed to be relevant for navigation in worms \cite{Ikeda2020-tw}, such as omega-shape and J-shape turns. Future work could explore whether adding these additional features to dPAW improves agreement. 

Another main assumption in dPAW, as well as other chemotaxis characterization in previous work \cite{Pierce-Shimomura1999-nt, Iino2009-al}, is that the behavioral responses are stationary through time. It is possible that worms can respond differently in response to different local concentration gradients. Future work is needed to characterize non-stationary behavioral dynamics during chemotaxis.

When investigating the relationship between the biased random walk (BRW) indices, the weathervaning  (WV) indices and the weighted chemotaxis index (wCI)  (Fig 6b), we were surprised and initially puzzled to find that changes to the indices did not always correspond to a change in the chemotaxis performance. In other words, chemotaxis performance, at least by this measure, isn't trivially the sum of the BRW and WV indices. For instance, AIA(-) and AIB(-) worms  are appetitive trained have the same weighted CI as wild-type, but their BRW and WV index are significantly altered. Previous work has also observed a similar mismatch between indices for behavioral strategies and chemotaxis performance \cite{Iino2009-al}.

Based on this observation, we speculate that it is possible that ablated or disrupted worms can adaptively employ different strategies, including those not captured by the BRW or WV indices, to reach similar chemotaxis performance. For instance, we qualitatively observed that AIB(-) worms have less sharp turns and have curving trajectories in space, but still end up moving up gradient (Fig. \ref{fig:SI10}), suggesting a possibly alternate strategy. In contrast, AIY(-) worms have high rate of sharp turns and move up gradient with non-smooth trajectories. Future work with models that have alternative strategies or more hierarchical structures are needed to characterize the complexity of these chemotaxis patterns.












\begin{figure}
\centering
\includegraphics[width=\textwidth]{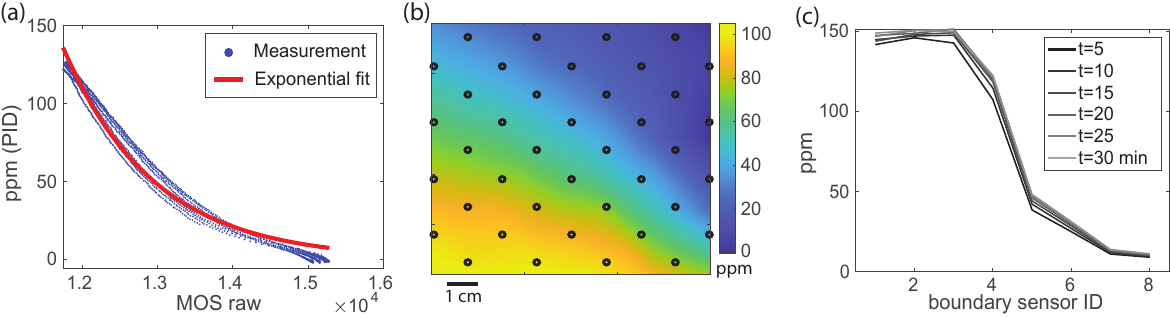}
\caption{Odor landscape experienced by the animal is known,  stable across space and time, and calibrated against a photoionization detector. \textbf{(a)} An array of metal-oxide sensors (MOS) is used to monitor odor concentration. The metal-oxide sensor is calibrated against a  a downstream photoionization detector (PID) to provide parts per million of butanone. \cite{Chen2023-fy}. \textbf{(b)} Odor gradient experienced by the animal in a typical experiment, as measured by the full array of sensors indicated in (black circles). Inferred odor concentration via interpolation is shown . \textbf{(c)} The odor concentration profile along the boundary is stable across the duration of the experiment. Readout from a row of boundary sensors downstream from the odor flow path during animal experiments.}
\label{fig:SI1}
\end{figure}

\begin{figure}
\centering
\includegraphics[width=.55\textwidth]{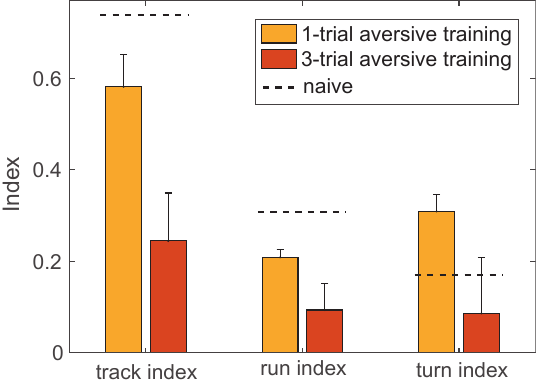}
\caption{Repeated aversive training strengthens the effect of learning. We compared conditions with single-trial of aversive training and with three repetitions. The chemotaxis performance and the usage of two behavioral strategies are computed with the same metrics shown in Figure 1. The values are compared to the mean of naive condition shown in dash line. Error bar shows standard error of mean from 7 plates.}
\label{fig:SI2}
\end{figure}

\begin{figure}
\centering
\includegraphics[width=.7\textwidth]{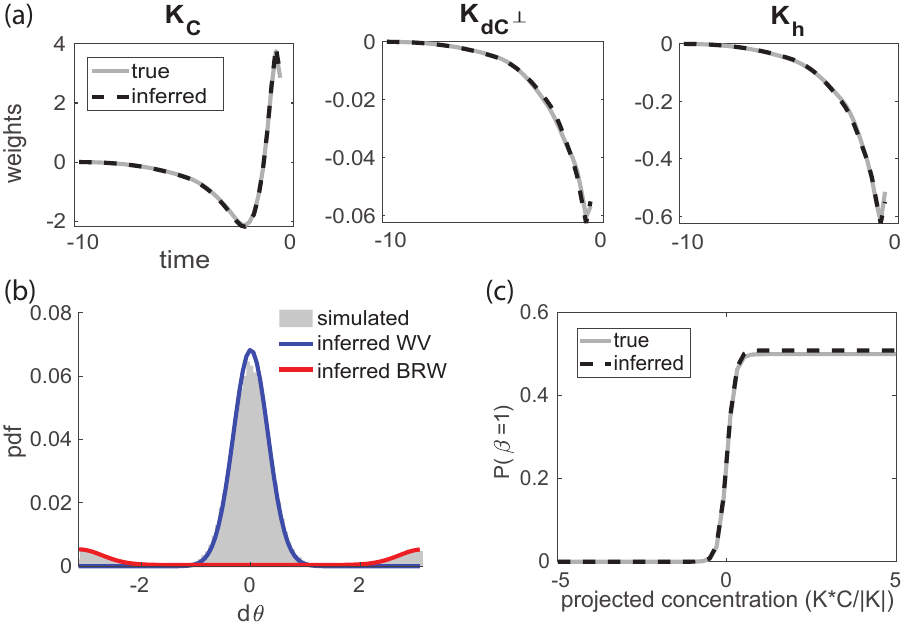}
\caption{The parameters that dPAW learns from the simulation agree with ground truth. Simulated time series is generated with known ground truth parameter, then used to train dPAW.  \textbf{(a)}  Inferred kernels agree with ground truth. \textbf{(b)} The $d\theta$ distribution simulated from ground-truth parameters overlaid with the inferred densities that come from the same generative model. Densities corresponding to weathervaning (WV) and biased random walk (BRW) strategies are shown in color. \textbf{(c)} The sigmoid-like function for pirouette decision as a function of concentration projected onto the normalized kernel $K_C$ is recovered from inference.}
\label{fig:SI3}
\end{figure}

\begin{figure}
\centering
\includegraphics[width=\textwidth]{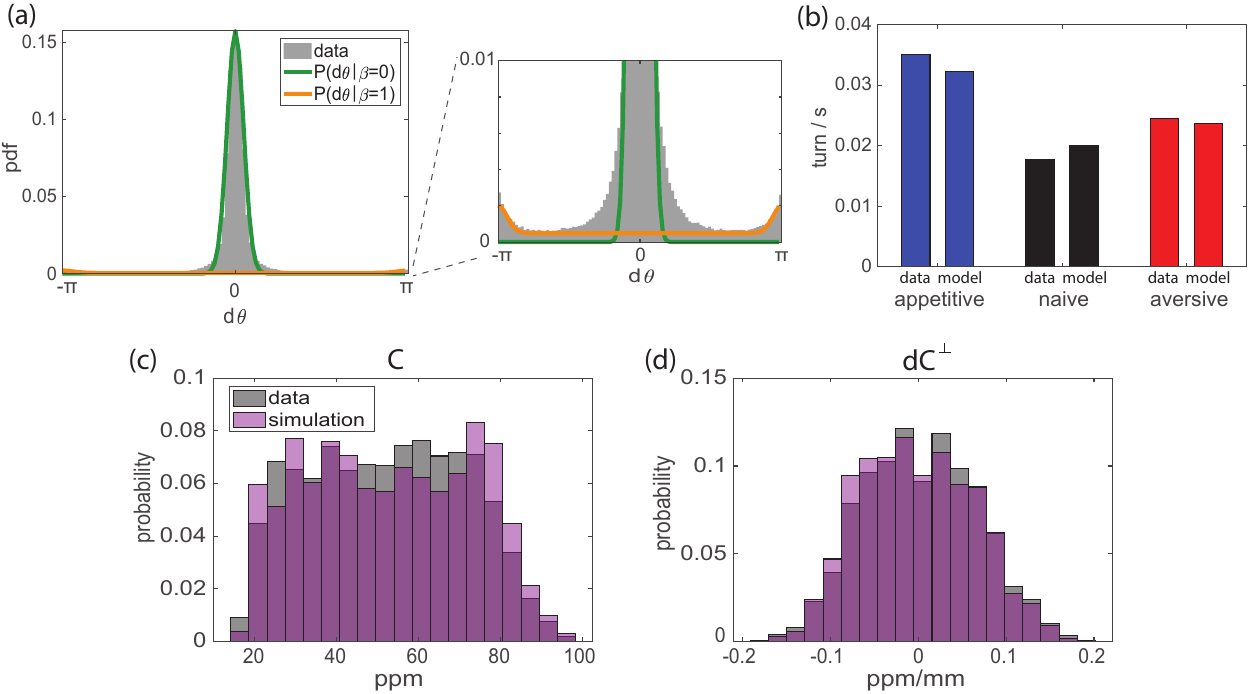}
\caption{The inferred dPAW parameters recapitulate measured sensory and behavioral statistics. \textbf{(a)} The experimentally observed $d\theta$ distribution overlaid with the inferred mixture of two strategies. Zoom-in for a lower probability regime is shown on the right to visualize the pirouette angles ($\beta=1$). \textbf{(b)} The empirical turn rate ($d\theta > 150$ degrees) in data and predicted by the inferred dPAW. The empirical and simulated concentration $C$ distribution \textbf{(c)} and perpendicular concentration difference $dC^{\perp}$ \textbf{(d)} are shown.}
\label{fig:SI4}
\end{figure}

\begin{figure}
\centering
\includegraphics[width=.9\textwidth]{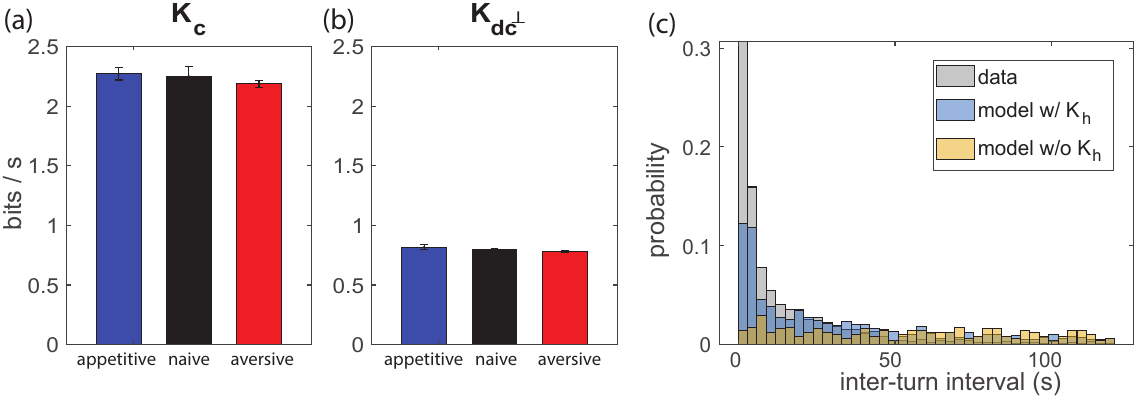}
\caption{The dPAW model has highest performance when it explicitly includes kernels for concentration, perpendicular concentration, and behavior history. Improvement in test log-likelihood (in units of bits / s) due to the inclusion of the specific dPAW model filters is shown. \textbf{(a)} Log-likelihood improvement of full dPAW model compared to the model with ablated kernel $K_C$. Error bar shows standard deviation across 7 fitted ensembles of trajectories. \textbf{(b)} Same as (a) but with ablated kernel $K_{dC^{\perp}}$. \textbf{(c)} The distribution of inter-turn-intervals observed from data, compared to trajectories simulated from the inferred dPAW with or without kernel $K_h$.
It has been reported that worms produce sharp turns in bouts with specific time scales, and the original definition of a pirouette included events defined by multiple sharp turns close together in time \cite{Pierce-Shimomura1999-nt}. The results shows that the dPAW model can generate tracks that have similar kinetics with the history kernel and without explicitly modeling state-transitions.
}
\label{fig:SI5}
\end{figure}

\begin{figure}
\centering
\includegraphics[width=.9\textwidth]{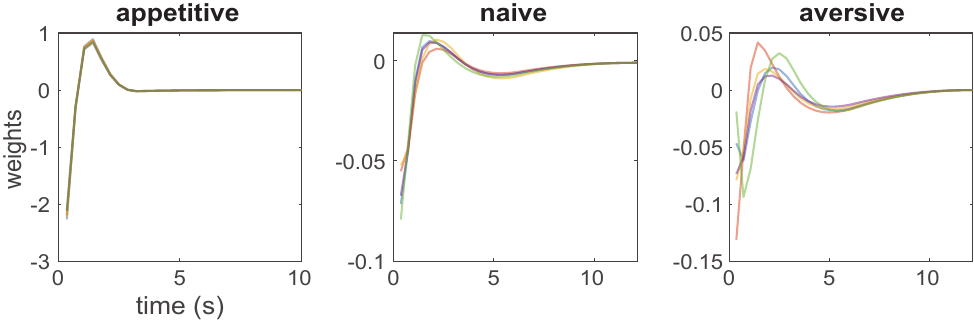}
\caption{Variability of the inferred kernels $K_C$ across 5 datasets subsampled (without replacement) from the full dataset, for each of the three training conditions. The inferred kernels $K_C$ for each subsample are shown in different color coded lines. The estimated kernels were highly consistent in appetitive condition, but more variable in naive and aversive conditions. 
}
\label{fig:SI6}
\end{figure}

\begin{figure}
\centering
\includegraphics[width=.6\textwidth]{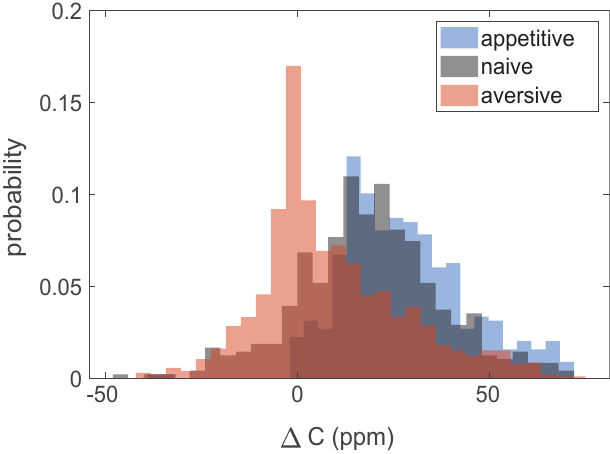}
\caption{Histogram of the odor concentration change $\Delta C$ along all trajectories observed after three different training conditions. The aversive condition has a qualitative shift away from appetitive and naive conditions. However, as quantified in Figure 4a, concentration alone has lower prediction power compared to the full dPAW model.}
\label{fig:SI7}
\end{figure}

\begin{figure}
\centering
\includegraphics[width=\textwidth]{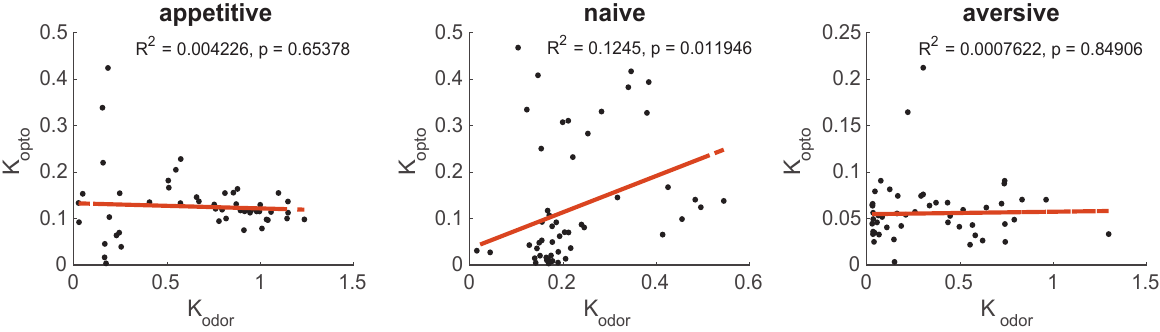}
\caption{Optogenetic kernel weights co-vary with odor kernel weights in naive but not learned conditions. 
To examine the relation between odor and optogenetic sensorimotor processing, we sub-sampled trajectories recorded from chemotaxis with optogenetic perturbation. For each subsampled navigation trajectories, we fit the statistical model for pirouette probability (equation 5) and computed the norm of kernels $K_{opto}$ and $K_{odor}$. We re-sampled ensemble of 100 trajectories from 300-1000 trajectories across three learning conditions. 
The scatter plots compare the norm of kernels $K_{opto}$ and $K_{odor}$ inferred from sub-sampled datasets.  The scattered samples show weak but significant correlation in naive condition. This means that subset of trajectories that vary and have less weight on odor signal would also have less optogenetic response. In contrast, aversive worm have clustered small weight on odor and is uncorrelated to the optogenetic input. Together, the sub-sampling result is consistent with the observation in Figure 5 and supports the finding that tracks going down gradient (less weight on odor input) also respond less to optogenetic input.}
\label{fig:SI8}
\end{figure}

\begin{figure}
\centering
\includegraphics[width=\textwidth]{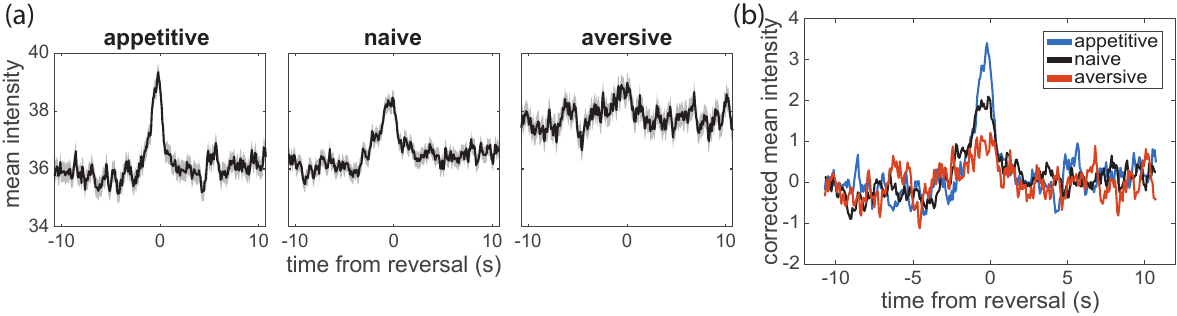}
\caption{Optogenetic kernels extracted from white-noise optogenetic stimulation measurements.  Changes to these optogenetic kernels mimic changes to the odor kernels that were observed in odor landscape. 
\textbf{(a)} Behavioral triggered average (BTA) for reversal computed from Gaussian white noise stimuli after three training conditions. The shaded area shows standard error of mean around the mean intensity ($\mu$W/mm$^2$). \textbf{(b)} Comparing BTA from (a) after baseline subtraction for three training conditions.
}
\label{fig:SI9}
\end{figure}

\begin{figure}
\centering
\includegraphics[width=.8\textwidth]{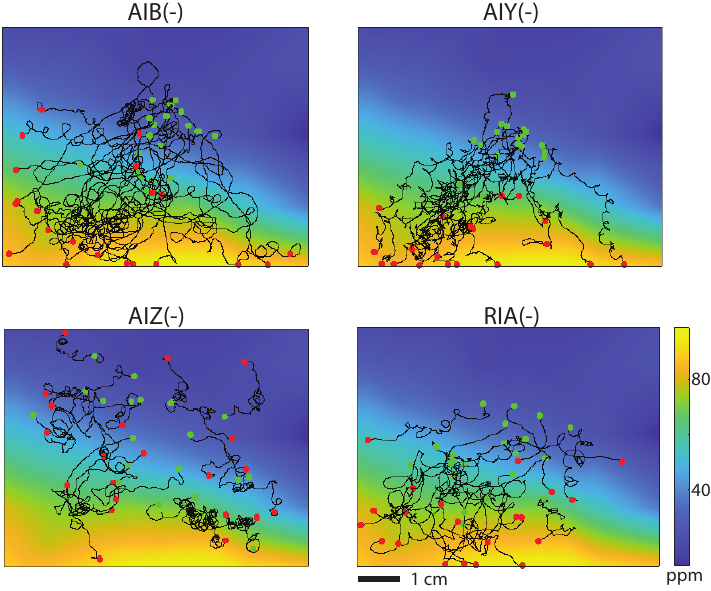}
\caption{Disruptions to different interneurons result in qualitatively different locomotion trajectories.  Trajectories form worms after appetitive training for disruptions to four different classes of interneurons are shown. Green dots and red dots indicate the starting point and ending point of each track, respectively.}
\label{fig:SI10}
\end{figure}

\begin{figure}
\centering
\includegraphics[width=.8\textwidth]{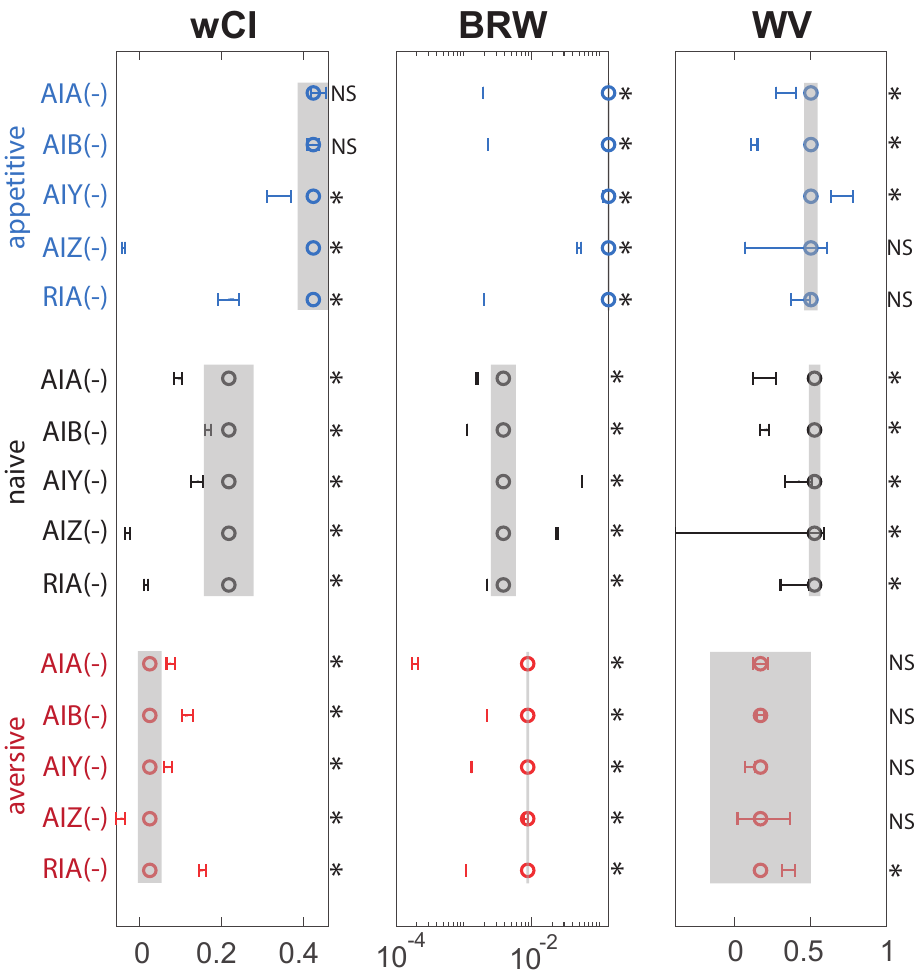}
\caption{Statistical tests for learned chemotaxis and behavioral strategies across worm strains with disrupted interneurons. For wild type worms, the mean values of weighted chemotaxis index (wCI), biased random walk (BRW), and weathervaning (WV), are shown in circle in each panel. The standard deviation of these values are indicated with the shaded area. Three different learning experiences are shown in color. 
For different mutant strains across different learning experiences, the mean and standard deviation are plotted for comparison to wild-type. The mean and standard deviation for wCI are computed from 20 re-sampled trajectories. The mean and standard deviation for BRW and WV are computed from 100 samples from the maximum likelihood estimation of kernels and the Hessian around it. We conduct t-test between the samples from mutant measurements and wild type measurements and indicate the result on the right side of each panel ($*$ for $p<0.01$ or non-significance, NS; using Bonferroni correction for multiple comparisons test).
}
\label{fig:SI11}
\end{figure}





\begin{table}\centering
\caption{Table for worm strains}

\begin{tabular}{lrrrr}

Strain & Designation & Genotype & Source or reference \\
\midrule
Wild-type \emph{C. elegans} & N2 & -- & CGC \\
ChR2 expressed in AWC$^{ON}$ & AML105 & \emph{wtfIs32 [str-2p::ChR2 (H134R)::GFP (50 ng/µl), myo-3p:mCherry (10 ng/µl)]}
& this work  \\
AIB(-) activated potassium channel & AML580 & \emph{wtfIs491 [inx-1p::twk-18(gf)::mcherry; unc-122p::rfp]}
& \cite{Chen2023-fy}\\
AIA(-) miniSOG & IK3240 & \emph{njls115[ins-1p::flp,gcy-28dp::frt::tomm-20(N'-55AA)::miniSOG,ges-1p::nls-GFP](IV)}
& \cite{Ikeda2020-tw} \\
AIY(-) miniSOG & IK2962 & \emph{njls87[AIYp::tomm-20(N'-55AA)::miniSOG,ges-1p::nls-GFP](IV)}
& \cite{Ikeda2020-tw} \\

AIZ(-) miniSOG & IK3241 & \emph{njls116[acc-2p::flp,odr-2(2b)p::frt::tomm-20(N'-55AA)::miniSOG,ges-1p::nls-GFP](IV)}
& \cite{Ikeda2020-tw} \\
RIA(-) miniSOG & IK3289 & \emph{njls123[glr-3p::tomm-20(N'-55AA)::miniSOG,ges-1p::nls-GFP](V)}
& \cite{Ikeda2020-tw} \\

\bottomrule
\end{tabular}
\end{table}

% ...
\begin{table}\centering
\caption{Table for experiments and figures}

\begin{tabular}{lrrrr}

Experiment & Strain & Odor & Optogenetics & Figures \\
\midrule
Odor navigation  & N2  &  \checkmark & x & Figure 1-4, Fig. S2-7 \\
Simultaneous odor and optogenetics   & AML105 & \checkmark & pulse & Figure 5, Fig. S8 \\
White noise optogenetics & AML105 & x & time-varying intensity white-noise & Fig. S9 \\
Interneuron disruption & AML580,IK3240,IK3241,IK3289,IK2962  & \checkmark & x & Figure 6, Fig. S10,11 \\

\bottomrule
\end{tabular}
\end{table}

\FloatBarrier




\dataset{Data}{Data to generate all figures in this work are posted on figshare website: \url{10.6084/m9.figshare.24764403}.}

\dataset{Code}{Code used for analysis and to generate figures are on github: \url{https://github.com/Kevin-Sean-Chen/Chemotaxis_function}. This repository would be updated upon publication.}

\bibliography{odor_SI_ref}